\documentclass[a4paper,fleqn,usenatbib]{mnras}

\usepackage{newtxtext,newtxmath}
\usepackage[T1]{fontenc}
\usepackage{ae,aecompl}

\usepackage[utf8]{inputenc} % If utf8 encoding
\usepackage{graphicx}	% Including figure files
\usepackage{amsmath}	% Advanced maths commands
\usepackage{amssymb}	% Extra maths symbols
\usepackage{subcaption} % kvuli subfigure
\usepackage{xcolor}

\title[Hot  atmospheres of rotating galaxies]{Hot gaseous atmospheres of rotating galaxies observed with {\it XMM-Newton}}

\author[A. Jur\'{a}\v{n}ov\'{a} et al.]{A. Jur\'{a}\v{n}ov\'{a}$^{1,2,3}$\thanks{E-mail: a.juranova@sron.nl},
	N. Werner$^{1,4,5}$,
	P. E. J. Nulsen$^{6,7}$,
	M. Gaspari$^{8,9}$\thanks{{\it Lyman Spitzer Jr.} Fellow},
	K. Lakhchaura$^{10}$,
    \newauthor R. E. A. Canning$^{11,12}$,
    M. Donahue$^{13}$,
    F. Hroch$^{1}$,
    G. M. Voit$^{13}$
	\\
	$^{1}$Department of Theoretical Physics and Astrophysics, Masaryk University, Kotl\'a\v{r}sk\'a 2, 611 37 Brno, Czech Republic\\
	$^{2}$SRON Netherlands Institute for Space Research, Sorbonnelaan 2, 3584 CA Utrecht, The Netherlands\\
	$^{3}$Anton Pannekoek Institute, University of Amsterdam, Postbus 94249, 1090 GE Amsterdam, The Netherlands\\
	$^{4}$School of Science, Hiroshima University, 1-3-1 Kagamiyama, Higashi-Hiroshima 739-8526, Japan\\
	$^{5}$MTA-E\"otv\"os Lor\'and University Lend\"ulet Hot Universe Research Group, H-1117 P\'azm\'any P\'eter s\'et\'any 1/A, Budapest, Hungary\\
	$^{6}$Harvard-Smithsonian Center for Astrophysics, 60 Garden Street, Cambridge, MA 02138, USA\\
	$^{7}$ICRAR, University of Western Australia, 35 Stirling Hwy, Crawley,	WA 6009, Australia \\
	$^{8}$Department of Astrophysical Sciences, Princeton University,  4 Ivy Lane, Princeton, NJ 08544-1001, USA\\
	$^{9}$INAF - Osservatorio di Astrofisica e Scienza dello Spazio, via P. Gobetti 93/3, I-40129 Bologna, Italy\\
	$^{10}$MTA-ELTE Astrophysics Research Group, H-1117 P\'azm\'any P\'eter s\'et\'any 1/A, Budapest, Hungary\\
	$^{11}$Kavli Institute for Particle Astrophysics and Cosmology, Stanford University, 452 Lomita Mall, Stanford, CA 94305-4085, USA\\
	$^{12}$Department of Physics, Stanford University, 382 Via Pueblo Mall, Stanford, CA 94305-4060, USA\\
	$^{13}$Physics \& Astronomy Department, Michigan State University, East Lansing, MI 48824-2320, USA
}
\date{Accepted XXX. Received YYY; in original form ZZZ}

\pubyear{2020}

\begin{document}
\label{firstpage}
\pagerange{\pageref{firstpage}--\pageref{lastpage}}
\maketitle

\begin{abstract}
X-ray emitting atmospheres of non-rotating early-type galaxies and their connection to central active galactic nuclei have been thoroughly studied over the years. However, in systems with significant angular momentum, processes of heating and cooling are likely to proceed differently. We present an analysis of the hot atmospheres of six lenticulars and a spiral galaxy to study the effects of angular momentum on the hot gas properties.
We find an alignment between the hot gas and the stellar distribution, with the ellipticity of the X-ray emission generally lower than that of the optical stellar emission, consistent with theoretical predictions for rotationally-supported hot atmospheres. The entropy profiles of NGC 4382 and the massive spiral galaxy NGC 1961 are significantly shallower than the entropy distribution in other galaxies, suggesting the presence of strong heating (via outflows or compressional) in the central regions of these systems. Finally, we investigate the thermal (in)stability of the hot atmospheres via criteria such as the TI- and $C$-ratio, and discuss the possibility that the discs of cold gas present in these objects have condensed out of the hot atmospheres. \end{abstract}


%%%%%%%%%%%%%%%%%%%%%%%%%%%%%%%%%%%%%%%%%%%%%%%%%%

%%%%%%%%%%%%%%%%% BODY OF PAPER %%%%%%%%%%%%%%%%%%

\section{Introduction}

Many massive early-type galaxies are permeated by hot gas, but its long-lasting presence, the role of the central supermassive black hole and the relation to colder gas phases have not been fully understood \citep[for a recent review see][]{werner2019}. Effects of significant net angular momentum of the X-ray atmospheres could be of considerable importance to the current view, affecting mostly relatively lower-mass host galaxies \citep{Eskridge1995,Sarzi2013,Nergi2014a,Negri2014b,Gaspari2015,defelippis2020}.

So far, most X-ray studies have focused on the hot atmospheres permeating massive slow rotating giant elliptical galaxies. However, around 80 per cent of the 260 early type galaxies in the  ATLAS$^{\rm 3D}$ sample are regular rotators \citep{Atlas3DII}, which often have flattened discy morphologies.
Interestingly, observations revealed that the X-ray luminosity, $L_{\rm X}$, of an early-type galaxy is related to its morphology, determined by its distribution of stars. Derived ratios of $L_{\rm X}/L_{\rm B}$ (to reduce the dependence on stellar mass) for `pure' ellipticals were found to be systematically higher than those of `discy' ellipticals and lenticulars \citep{Eskridge1995,Sarzi2013}.

The flattened shape of the gravitational potential and rotational support, which has been confirmed for flattened early-type galaxies \citep{Atlas3DIII}, should allow easier development of outflows. Furthermore, it has been shown by \citet{Pellegrini2012a} that in low-mass systems, the energy injection through supernovae can become dominant and thus have a significant effect on the properties of their X-ray haloes \citep[see also e.g.][]{Davids1990, Mathews2003}.

According to both observations and theoretical predictions, cool and cold gas in brightest cluster galaxies is thought to be a product of cooling from the hot intracluster medium \citep[e.g.][for a recent review]{johnstone1987,heckman1989,Bridges1998,Conselice2001,McDonald2011, Gaspari2020} and partly a product of stellar mass loss \citep{Voit2011}. About half of the nearby X-ray bright giant elliptical galaxies contain H$\alpha$+[NII] emitting gas  that most likely cooled from the hot X-ray atmospheres \citep{Lakhchaura2018}.  Recent surveys focusing on CO emission in early-type galaxies \citep{Combes2007, Davis2019, Young2011} found the presence of molecular gas in approximately 25 per cent of studied objects. 
For the ATLAS$^{\rm 3D}$ sample, \cite{Young2011} found no correlation of the molecular gas mass with stellar (K-band) luminosity, which implies that the molecular gas does not originate purely in stellar mass loss.
\citet{Babyk2019} found a correlation between the mass of the X-ray atmosphere and molecular hydrogen content. Molecular hydrogen was reported in systems with cooling times as short as $10^9~ \rm yr$  \citep[e.g.][]{Temi2018}, indicating the conditions are right for condensation of cool, dense gas from hot atmospheres. 

Hydrodynamic simulations of rotationally supported systems predict that cooling from the hot phase should lead to the formation of multiphase discs \citep{Gaspari2015}. Recent \textit{XMM-Newton} observations of the rotating lenticular galaxy, NGC~7049, support this proposition. In the otherwise isothermal gas, spectral signatures of ongoing cooling were detected in a region co-spatial with the plane of rotation of the galaxy as well as a multiphase disc ranging from warm gas to a cold molecular phase \citep{Juranova2019}.

Here, we focus on the properties of a sample of hot atmospheres permeating six rotationally-supported lenticular galaxies and a massive spiral galaxy and compare our findings with non-rotating systems.
The paper is structured as follows. In Sect. \ref{sec:analysis}, we present the selected objects and describe the analysis of the \textit{XMM-Newton} EPIC observations used. The main results are shown in Sect. \ref{sec:results}, specifically focusing on the morphology of the hot gas (\ref{sec:imaging}) and its thermodynamic properties (\ref{sec:thermodynamics}), and we address the thermal stability of the hot atmospheres (\ref{sec:stability}). In Sect. \ref{sec:discussion}, we discuss our findings and present our conclusions in Sect. \ref{sec:conclusions}.

\section{The sample and data reduction}\label{sec:analysis}

\subsection{The sample}

To study the properties of the hot gas in rotating  galaxies, we searched the volume-limited ATLAS$^{\rm 3D}$ sample \citep{Atlas3DI} for systems where the 2D maps of line-of-sight stellar velocities indicate ordered rotation \citep[classified as `regular rotators' in][]{Atlas3DII} and the {\it XMM-Newton} archive contains sufficient data to determine the gas density and temperature in at least three concentric annular regions. All objects selected in our sample have the ratio of rotational velocity to velocity dispersion greater than 1/3 (see Table \ref{tab:geometry}).
All galaxies in the ATLAS$^{\rm 3D}$ sample are located at a distance $ D<42~\mathrm{Mpc} $, which allows us to perform a spatially resolved analysis of their extended X-ray emitting atmospheres with {\it XMM-Newton}. The sensitivity of {\it XMM-Newton} in the 0.3--2~keV band, where most of the X-rays emitted by S0 galaxies emerge, makes this satellite uniquely suited for the study of these systems.

\begin{figure}
	\centering
	\begin{subfigure}[b]{.32\linewidth}
		\includegraphics[width=\linewidth]{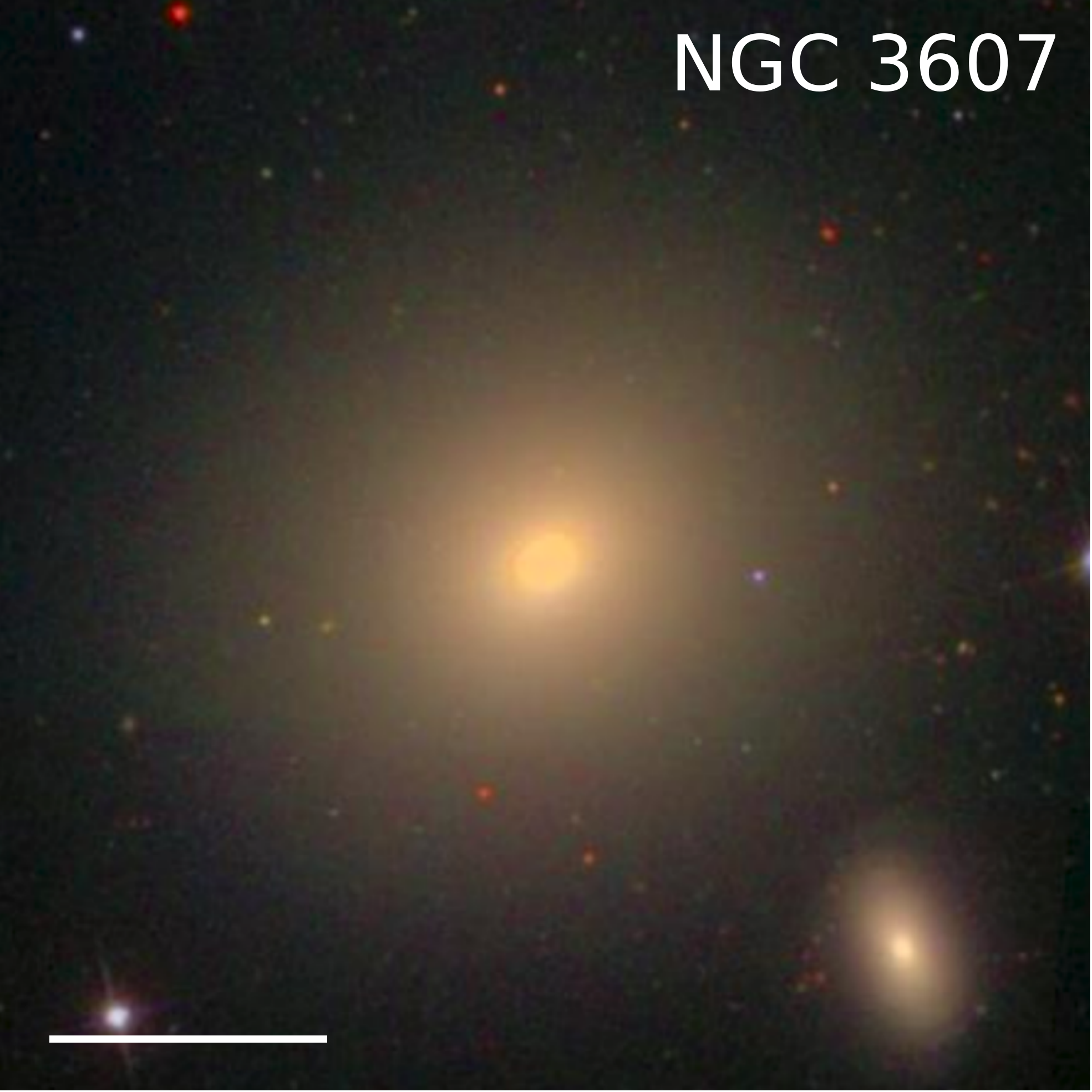}
	\end{subfigure}
	\begin{subfigure}[b]{.32\linewidth}
		\includegraphics[width=\linewidth]{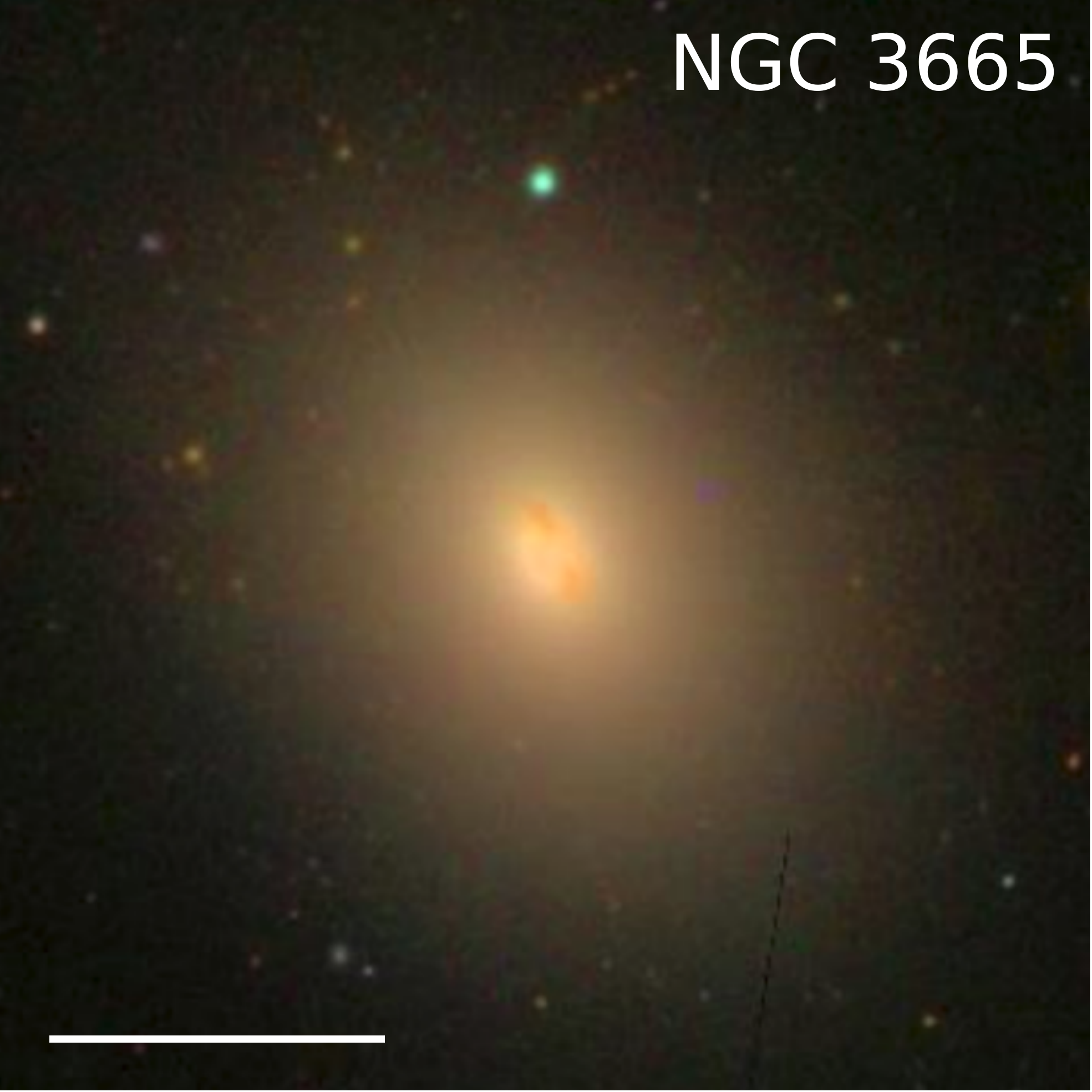}
	\end{subfigure}\vspace{0.03cm}
	\begin{subfigure}[b]{.32\linewidth}
		\includegraphics[width=\linewidth]{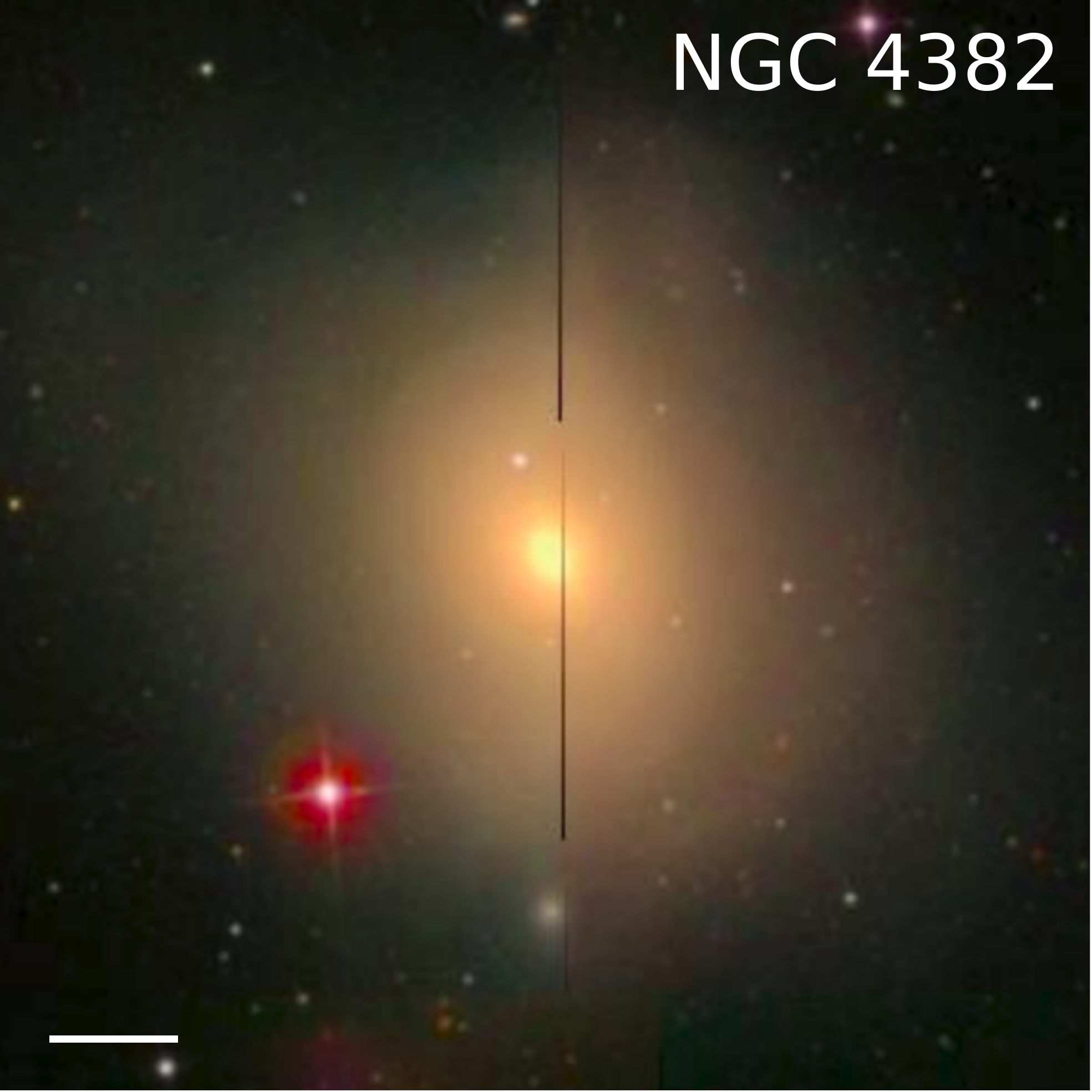}
	\end{subfigure}

	\begin{subfigure}[b]{.32\linewidth}
		\includegraphics[width=\linewidth]{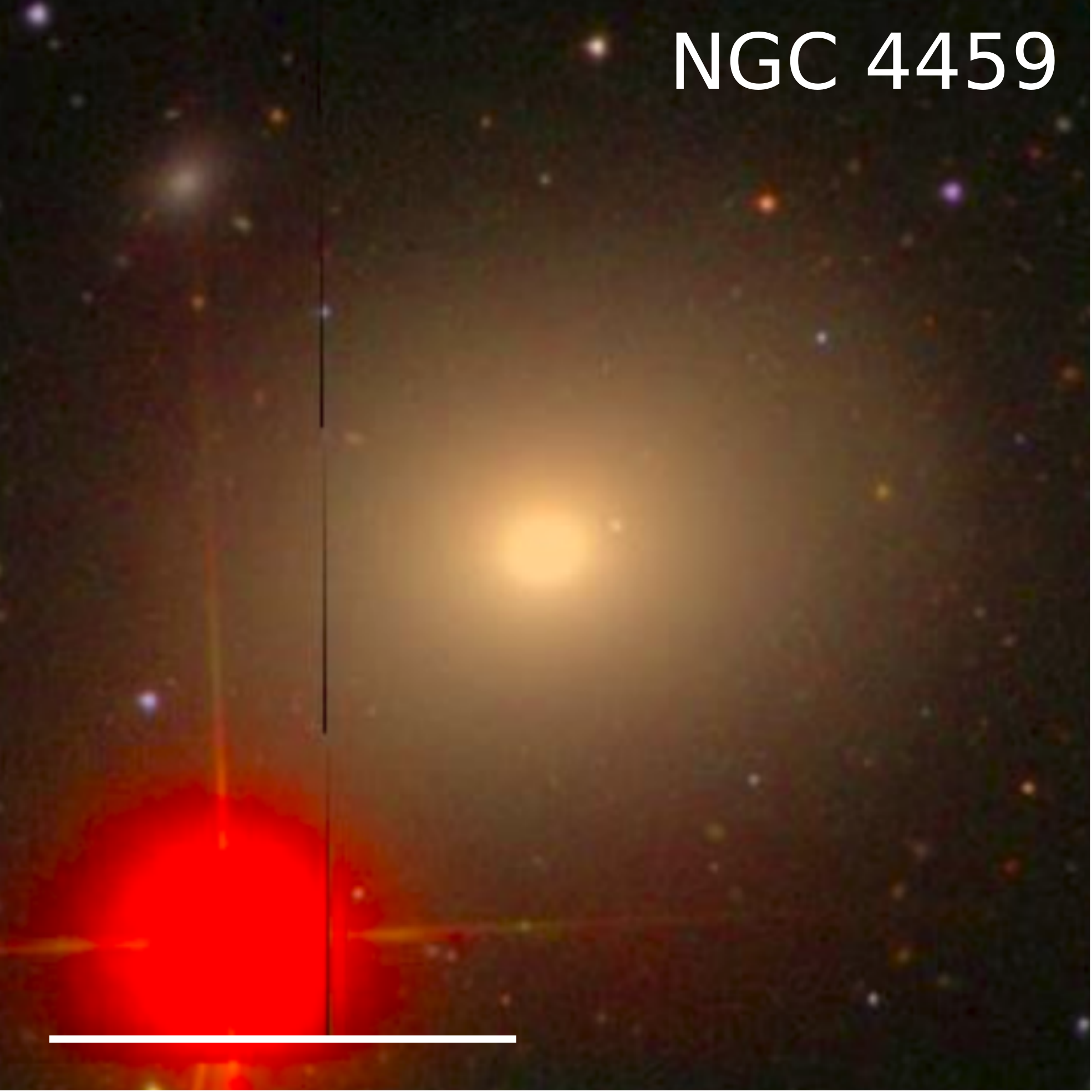}
	\end{subfigure}
	\begin{subfigure}[b]{.32\linewidth}
		\includegraphics[width=\linewidth]{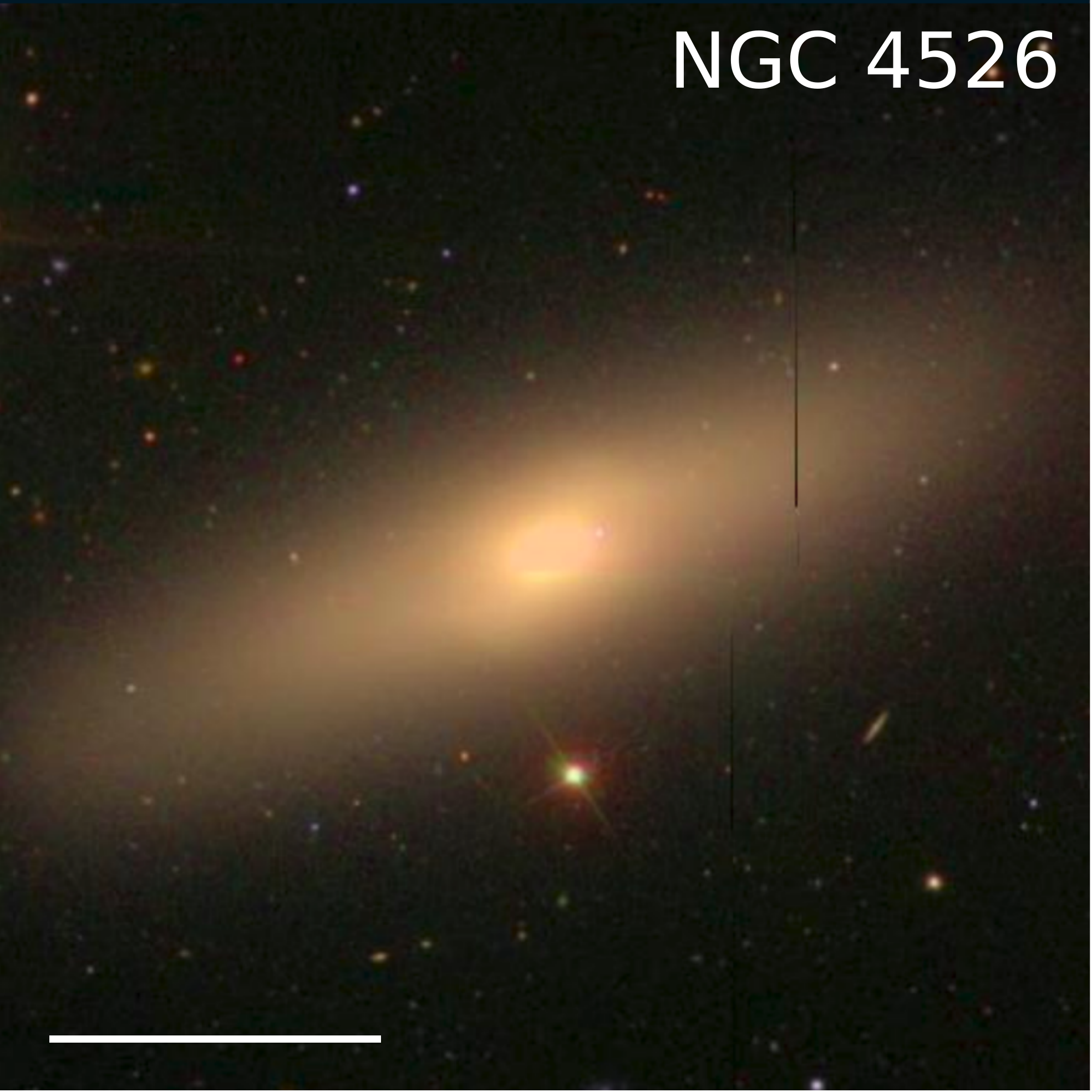}
	\end{subfigure}\vspace{0.03cm}
	\begin{subfigure}[b]{.32\linewidth}
		\includegraphics[width=\linewidth]{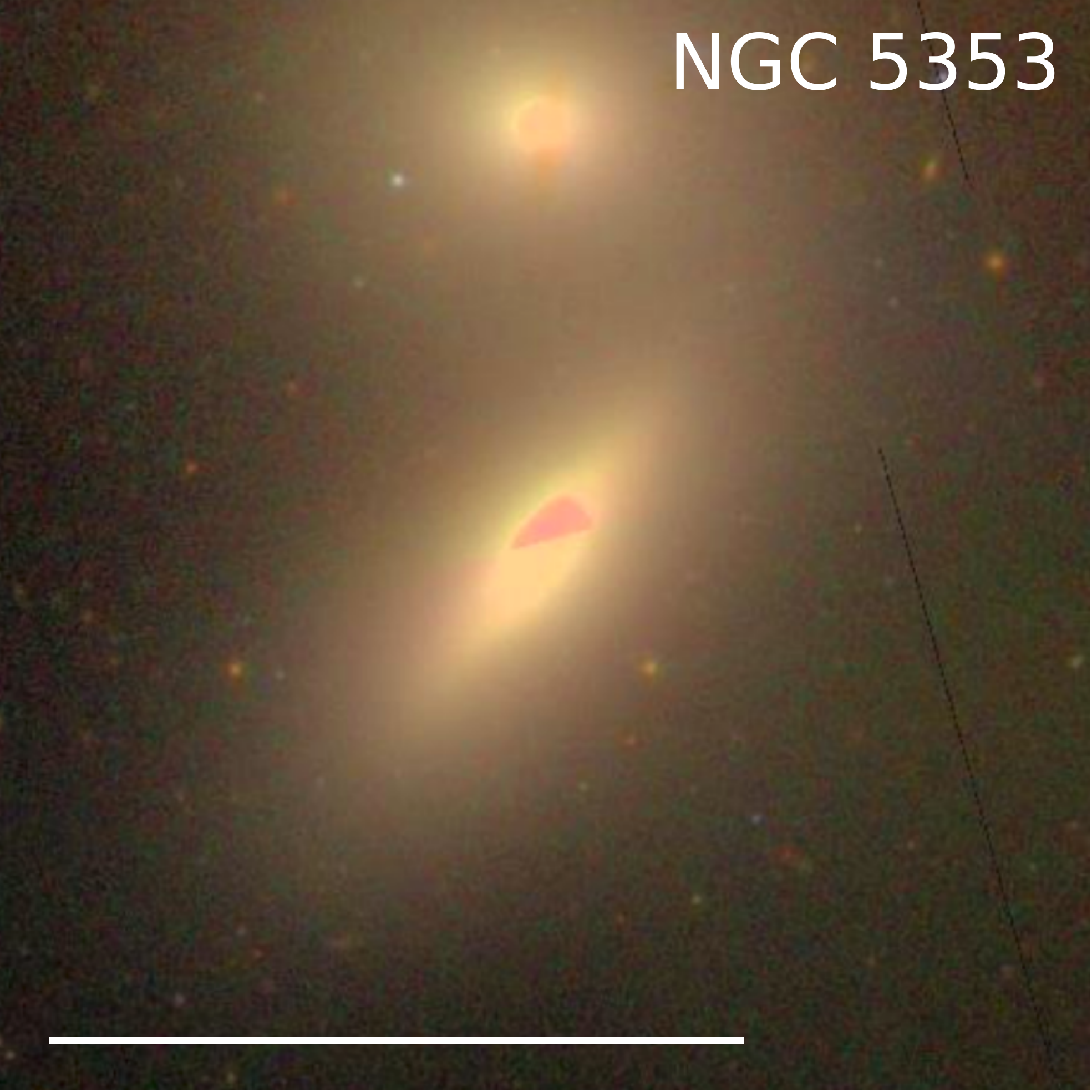}
	\end{subfigure}

	\begin{subfigure}[b]{.32\linewidth}
		\includegraphics[width=\linewidth]{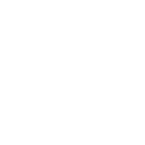}
	\end{subfigure}
	\begin{subfigure}[b]{.32\linewidth}
		\includegraphics[width=\linewidth]{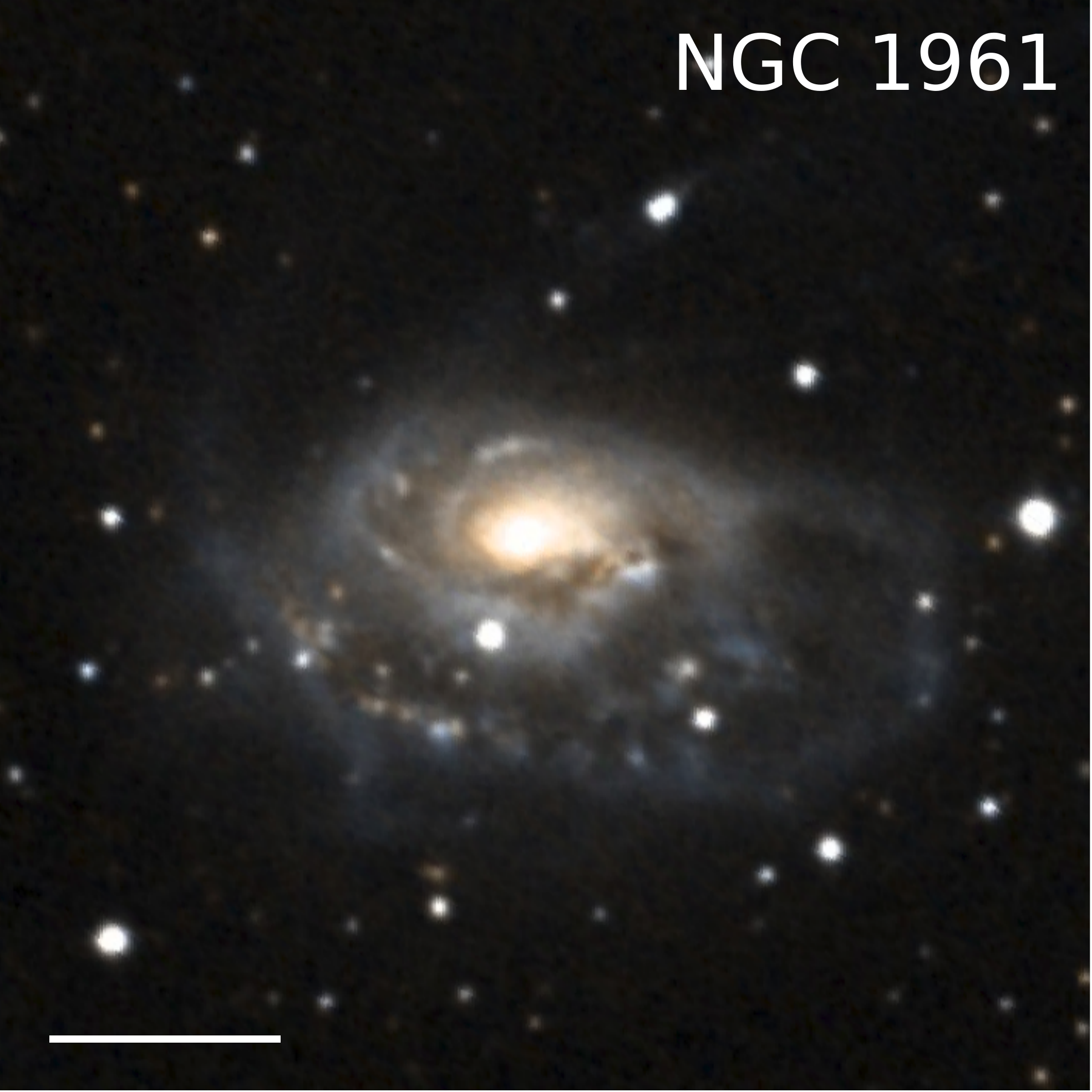}
	\end{subfigure}
	\begin{subfigure}[b]{.32\linewidth}
		\includegraphics[width=\linewidth]{empty-cut.png}

	\end{subfigure}
		\caption{Optical images of the studied sample of S0 galaxies and NGC~1961. The images are taken from Sloan Digital Sky Survey (SDSS, data release 7) with the exception of NGC~1961, which is from Digitized Sky Survey 2. The SDSS images are compositions of \textit{g}, \textit{r}, and \textit{i} \citep{Smith2002} imaging data.
		The solid line in the lower-left corner of every image represents a scale of 10 kpc.}
		\label{fig:optical}
\end{figure}

The selection resulted in six S0 galaxies and the X-ray bright spiral galaxy NGC~1961. Optical images of the whole sample are shown in Fig. \ref{fig:optical}. The most important properties of these galaxies are listed in Tables~\ref{tab:geometry} and \ref{tab:content}. Their available {\it XMM-Newton} observations, denoted by observation IDs (OBSIDs), are listed in Table~\ref{tab:data}. 

	\subsubsection{NGC~3607}\label{NGC3607}
	NGC~3607 is an unbarred lenticular galaxy (morphological type SA0) and the brightest of the 18 member galaxies of the Leo II group \citep{Giuricin2000}. It has the highest star formation rate (SFR) in our sample, $ 0.42~\mathrm{M_{\odot}yr^{-1}} $ \citep{Amblard2014}. The presence of a rotating disc of cold molecular gas in NGC~3607 was confirmed from measurements of CO emission \citep{Alatalo2013}. \textit{Chandra X-ray Observatory} measurements do not show evidence for an X-ray bright AGN.

	\subsubsection{NGC~3665}\label{NGC3665}
	Another SA0 type galaxy in our sample, NGC~3665, is the brightest member of a group of 11 galaxies \citep{Makarov2011}. Of our sample, it has the highest luminosity emerging as emission features of polycyclic aromatic hydrocarbons \citep[PAH;][]{Kokusho2017},	triggered by absorption of far-ultraviolet light indicating a relatively high SFR. However, the SFR measured by other methods \citep{Amblard2014} in this system is still relatively low, at $\sim0.1~\rm M_{\odot}~yr^{-1}$. A rotating cold molecular disc reaching out to $\sim 3$~kpc from the centre of the galaxy was detected by \citet{Alatalo2013}. NGC~3665 is among our radio brightest S0s at 1.4 GHz and is the only one in our sample that shows a pronounced radio activity associated with an AGN. Twin jets were observed by the Very Large Array (VLA) \citep{Parma1986}, as well as core radio emission at $ 5~\rm GHz $ \citep{Liuzzo2009}. The mass of the supermassive black hole was measured by \citet{Onishi2017} to be $M_{\bullet} = 5.8 \times 10^8~\rm M_{\odot}$.

	\subsubsection{NGC~4382}\label{NGC4382}
	This object, also known as M~85, resides in the Virgo cluster. Its distance to the cluster centre is, however, almost 6 degrees, which is well outside the virial radius of Virgo. It is the only galaxy in our sample, which does not have a disc of cold gas near its centre. The morphological type of this galaxy is SA0 and the galaxy also shows shell-like structures. These features are formed by the stars of smaller infalling galaxies, due to the combined effects of Liouville’s theorem and phase wrapping \citep{Quinn1984}. The lack of cold gas is consistent with the observed low SFR of $0.002~\rm M_{\odot}~yr^{-1}$ \citep{Amblard2014}.
	
	\citet{Capetti2009} found that the core of NGC~4382 does not show any radio emission related to a central black hole. The current scaling $M_{\bullet}-L$ \citep{Kormendy1993} predict a black hole mass of $M_{\bullet} = 1.7 \times 10^9~ \rm M_{\odot}$, an estimate recently published by \citet{Graham2019}, who also identified an X-ray point source ($ L_{\rm X, AGN} = 8\times 10^{38}~\rm{erg~ s^{-1}}$) revealed by \textit{Chandra} observations as a counterpart to the central black hole.
	
	\subsubsection{NGC~4459}\label{NGC4459}
	Another member of the Virgo cluster, NGC~4459, is an unbarred lenticular galaxy with a dusty disc extending out to $r \sim 0.7~\rm kpc$, with observed blue clumps suggesting the presence of newborn stars \citep{Ferrarese2006}. Unable to detect any neutral hydrogen in their observations, \citet{Lucero2013} provide only an upper limit on its mass, $M_{\rm H\textsc{i}} < 1.7 \times 10^7~ \rm M_{\odot}$, despite the clear presence of molecular hydrogen (see Table~\ref{tab:content}), yielding a ratio $M_{\rm H_2}/M_{\rm H\textsc{i}} >21$.  Far- and mid-infrared emission of gas was studied in detail by \citet{Young2009} and led to a detection of $24\umu \rm m$ emission from a disc reaching out to $r\sim 2.6~ \rm kpc$, which exceeds almost four times the radius of the dusty structures and thus cannot be attributed to ongoing star formation. They also find that NGC~4459 has a so-called FIR-excess, i.e. FIR-to-radio flux density ratio exceeding a value of 3.04, a rare feature defined from observations of a large sample of galaxies in \citet{Yun2001}. In the centre of NGC~4459, \citet{Gavazzi2018} report a detection of $\rm H\,\alpha$ emission.
	An X-ray image from \textit{Chandra} revealed a point source identified as the central AGN \citep[$ L_{\rm X, AGN} = 10^{39} ~\rm{erg~ s^{-1}}$;][]{Gallo2010}.

	\subsubsection{NGC~4526}\label{NGC4526}
	The only barred lenticular galaxy (SAB) in our sample, NGC~4526, is also a member of the Virgo Cluster, but again outside the cluster virial radius. Having two entries in the New General Catalogue, it is also known as NGC~4560. Hereafter, we will only use the designation NGC~4526 for this object, which is also preferred in the scientific community. It is oriented nearly edge-on and also possesses a dusty disc in the plane of rotation, spanning over the central $r\sim 1.2~\rm kpc$. The SFR in NGC~4526 is the smallest among galaxies with a dusty disc in our sample \citep{Amblard2014} and the PAH luminosity the second highest, $\sim\!17\times 10^{41}~\rm erg\,s^{-1}$ \citep{Kokusho2017}. As in the case of NGC~4459, \citet{Lucero2013} did not detect neutral hydrogen and only placed the upper limit of $M_{\rm H\,\textsc{i}}< 1.9\times 10^7~\rm M_{\odot} $, while the total mass of $\rm H_2$ was measured well (see Table~\ref{tab:content}). The lower limit of their ratio is thus even larger than in NGC~4459: $M_{\rm H_2}/M_{\rm H\,\textsc{i}} >100$. Focusing on the warm gas content, \citet{Gavazzi2018} found $\rm H\,\alpha$ emission with a disc-like morphology in this galaxy. \citet{Davis2013} measured the mass of the central supermassive black hole to be $M_{\bullet} = 4.5 \times 10^8~ \rm M_{\odot}$.

	\subsubsection{NGC~5353}\label{NGC5353}
	This almost edge-on galaxy is a member of the compact group, HCG~68, with NGC~5350, NGC~5354 and two more objects \citep{Hickson1982}, accompanied by another 45 fainter galaxies and more member candidates \citep{Tully2008}. It has an effective radius of only $\sim3.4~\rm kpc$, which is smaller than any of the other 5 lenticular galaxies in our study. \citet{OSullivan2018} found a CO disc with a radius of $0.8~\rm kpc$ accompanied by a dusty disc in the central $r \sim 0.5~\rm kpc$ \citep{Goullaud2018}. This object is classified as a low-ionization nuclear emission-line region (LINER) galaxy \citep[e.g.][]{Saikia2018}. \citet{Sanchez2006} reported an X-ray bright nuclear point source spatially coincident with a radio source, which can be associated with the central AGN.

	\subsubsection{NGC~1961}\label{NGC1961}
	This rotation-dominated spiral galaxy is the brightest member of a small group of seven galaxies \citep{Tempel2016} and it has been classified as a LINER \citep{Carrillo1999}. Measurements of rotational velocity at distances exceeding $10~\rm kpc$ from the galaxy centre \citep{Rubin1979} have revealed that this spiral galaxy is exceptionally massive, and X-ray observations showing a point source located in the centre of the galaxy confirm the presence of an AGN \citep{Roberts2000}. The total mass of the galaxy is more than sufficient to retain its hot gaseous atmosphere, which has been observed by both \textit{Chandra} and \textit{XMM-Newton} \citep{Anderson2011, Bogdan2013}. \citet{Bogdan2013} compared available observations of NGC~1961 with results from numerical simulations finding that the galaxy is dark matter dominated -- baryons contribute only $11~\%$ of the total mass. In radio observations at $6$ and $18~\rm cm$ presented in \citet{Krips2007}, nuclear emission is accompanied by a $\sim 2\sigma$ signal resembling radio jets. Additionally, distorted $\rm H\textsc{i}$ morphology revealed that the gas is being stripped by the surrounding intragroup medium \citep{Shostak1982}. 
		
	\begin{table*}
		\caption{Basic observational properties of the studied sample. Distance $d$ is adopted from the ATLAS$^{\rm 3D}$ Project and references therein with the exception of NGC~1961, where the median value of distances from NED was taken, as well as values for redshift ($z$) in the second column. Where given, the values of $M_{200}$ were determined from globular clusters kinematics or the total mass of galaxy's globular clusters and published by \citet{kim2019} except for NGC~1961, where it was determined from the maximal stellar rotation velocity by \citet{Bogdan2013}. The effective radii $ R_{\mathrm{e}} $ and stellar velocity dispersions and rotational velocities are adopted from \citet{Atlas3DI}.}
	\begin{tabular}{cccccccc}
		\hline
		object& $ d $ & $ z $ & scale & $ R_{\mathrm{e}} $ & $\log M_{200}$ & $ \sigma_{v,\star} $ & $ v_{\mathrm{rot}, \star} $ \\
		NGC& Mpc & $ 10^{-3} $& $\rm arcsec~kpc^{-1}$ & $\mathrm{kpc}$ & $\mathrm{M_{\odot}}$ & $ \rm km\,s^{-1} $ & $ \rm km\,s^{-1}$ \\
		\hline
		3607& 22.2 & 3.14 & 9.29 & 4.24 & 12.85 & $222.0\pm4.0$ & $ 110.0\pm9.0 $ \\
		3665& 33.1 & 6.90 & 6.23 & 3.43 & $-$ & $215.0\pm8.5$ & $ 94.3\pm21.5 $ \\
		4382& 17.9 & 2.43 & 11.52 & 9.93 & 13.13 & $68.3\pm16.5$ & $ 176.0\pm3.5 $ \\
		4459& 16.1 & 3.98 & 12.81 & 2.82 & 12.67 & $75.0\pm20.0$ & $ 171.8\pm4.8 $ \\
		4526& 16.4 & 2.06 & 12.58 & 3.66 & 12.87 & $150.4\pm8.6$ & $ 246.0\pm6.0 $ \\
		5353& 35.2 & 7.75 & 5.86 & 2.58 & $-$ & $298.0\pm9.0$ & $ 284.0\pm4.8 $ \\
		\hline
		1961& 32.4 & 13.12 & 6.37 & 8.06 & 13.08 & $242.0\pm12.0$ & $ 326.8\pm9.5 $ \\
		\hline
	\end{tabular}
	\label{tab:geometry}
	\end{table*}

	\begin{table*}
	\caption{Multi-wavelength properties of the  analysed sample. $B$-band stellar luminosities $L_{B}$ and $B\!-\!V$ colour indexes are taken from HyperLEDA \citep{HyperLEDA}, $1.4~\rm GHz$ radio powers from \citet{Brown2011} and \citet{Condon2002}, PAH luminosities from \citet{Kokusho2017} and \citet{Stierwalt2014} in the case of NGC~1961, masses of molecular hydrogen from \citet{Young2011} and \citet{Combes2009} for NGC~1961, masses of atomic hydrogen from \citet{Young2014}, \citet{Haynes2018} for NGC~4526 and \citet{Haan2008} for NGC~1961. Star formation rates are taken from \citet{Amblard2014} and \citet{Davis2014}.}
	\begin{tabular}{ccccccccc}
		\hline
		object& $L_{B}$ &$B\!-\!V$& $ \log P_{\rm radio}$ & $L_{\rm PAH}$ & $ \log M_{\mathrm{H_2}} $ & $ \log M_{\mathrm{H\textsc{i}}} $ & SFR  \\
		NGC& $10^{10}\,\mathrm{L}_{B, \odot}$ &$\rm mag$& $ \rm W\,Hz^{-1} $& $ 10^{41}\,\rm erg\,s^{-1} $ & $ \rm M_{\odot} $ & $ \rm M_{\odot} $ & $ \mathrm{M_{\odot}\,yr^{-1}}$ \\
		\hline
		3607& $ 3.70 $ &$0.93$& $20.63$ & $7.8\pm6.2$ & $8.42$ & $<6.53$ & $ 0.420 $\\
		3665& $ 3.37 $ &$0.93$& $22.04$ & $45.1\pm9.9$ & $8.91$ & $<7.05$ & $ 0.109 $\\
		4382& $ 5.86 $ &$0.89$& $<19.79$ & $0.3\pm 5.5$ & $< 7.39$ & $<6.59$ & $ 0.002 $\\
		4459& $ 1.45 $ &$0.97$& $<19.63$ & $7.1\pm2.9$ & $8.24$ & $<6.53$ & $ 0.071 $\\
		4526& $ 2.42 $ &$0.98$& $20.61$ & $17.1\pm4.4$ & $8.59$ & $7.15$ & $ 0.028 $\\
		5353& $ 3.56 $ &$1.03$& $21.62$ & $2.3\pm6.7$ & $< 7.44$ & $<7.07$ & $ 0.095 $\\
		\hline
		1961& $ 22.91 $ &$0.86$& $22.82$ & $37.8\pm3.1$ & $10.39$ & $10.67$ & $ 9.24 $\\
		\hline
	\end{tabular}
	\label{tab:content}
	\end{table*}

\subsection{Data reduction}

The data were reduced with the standard procedures of the \textit{XMM-Newton} Science Analysis System version 17.0.0. Event lists from raw-data files were obtained using tasks \texttt{emchain} for EPIC-MOS and \texttt{epchain} for EPIC-pn. For EPIC-pn, we also created models of events received during readout (so-called out-of-time events). For this step, the \texttt{epchain} task was run for a second time to obtain the out-of-time (OOT) event list, which was, after proper scaling, subtracted from the pn events. Parts of the observations are affected by so called soft-proton flares. To filter out the affected time periods, we excluded the data where the count rate deviated from the mean by more than 1.5$\sigma$. The total and the filtered net exposure times, $ t_{\rm tot} $ and $ t_{\rm net} $, respectively, are given in Table~\ref{tab:data}. 

The extraction of images was performed using the ESAS package \citep{ESAS} in the 0.3--2.0 keV band. 

To study the thermodynamic properties of the X-ray emitting gas, we extracted spectra from several concentric annuli. If the number of photons was sufficiently high, the width of the annuli was determined by the angular resolution of \textit{XMM-Newton}. For the innermost annuli, the width was typically 15 arcsec ($\sim 1.2 \times$ the FWHM of the EPIC-pn or $\sim 3.4 \times$ the FWHM of the EPIC-MOS).

Typically, X-ray point sources of various origin are projected onto the extended emission of interest. To avoid their undesired contribution to the spectra, we encircled the emission of each point source and excluded the selected events during the spectral extraction procedure. As none of the X-ray atmospheres of interest covered the entire field of view, it was possible to create a spectrum of local background to be subtracted prior to spectral analysis.

\subsection{Spectral analysis}\label{subsec:analysis}
The spectra are fitted in the $0.5-5.0~\rm keV$ range.  Events in the $1.38-1.60~\rm keV$ range were excluded due to contamination by instrumental lines \citep[e.g.][]{Carter2007}.

In the following sections, we present results from both projected and deprojected spectra, focusing primarily on the latter.
For the analysis of projected spectra, we used the SPEX spectral fitting package \citep[][]{SPEX} version 3.04.00, which uses an extensive atomic database SPEXACT (version 2.07.00). SPEX allows fitting using the C-statistic \citep{Cash1979}, defined for Poissonian distribution of data. This is particularly useful for spectra with a low number of counts in a large set of bins, because the use of C-statistic does not require extensive binning. The data were binned to contain at least one count per bin. 
The spectra were deprojected using the so-called Direct X-ray Spectral Deprojection \citep[DSDEPROJ,][]{Russell2008}. Using this method, the spectra extracted from individual annuli are properly subtracted from those lying closer to their common centre. Unfortunately, the deprojected spectra could not be fitted using C-statistic, so $\chi^2$ minimisation was used instead. These spectra are binned to at least 25 counts per bin.

As the X-ray emitting gas can be described as a dilute plasma in collisional ionization equilibrium (CIE), we used the corresponding model \texttt{cie} in SPEX. At the energy resolution of EPIC, the spectral properties are set mainly by the gas temperature, emission measure, which translates to a normalisation of the spectrum, and elemental abundances. As the latter were not well-constrained by the spectral fits, we assumed the Solar relative abundances of \citet{Lodders2009} and fixed the metallicity to $0.5~\rm Z_{\odot}$.
We note that a bias in the overall metallicity would affect the derived physical quantities as follows: a factor of two difference in the measured and actual metallicity would result in 25 per cent bias in the gas density, and 17 per cent in the entropy. Slopes of radial profiles would be altered by less than 10 per cent in the presence of metallicity gradients \citep{werner2012}.

The fact that the gas temperature is not constant over the investigated volume of the atmosphere was accounted for by the parameter \texttt{sig} in the \texttt{cie} model. When non-zero, this parameter changes the model from single- to multi-temperature with a Gaussian distribution and the root-mean-square width equal to \texttt{sig}. Even though the assumption of a Gaussian temperature distribution does not necessarily reflect the true temperature distribution accurately, it still serves as a good approximation and adds only one free parameter to our model. The parameter \texttt{sig} was left free during fitting when a single-temperature model did not provide a good fit.

The unresolved population of low-mass X-ray binaries (LMXBs) was modelled with a power-law component with spectral index 1.6, in accordance with \citet{Irwin2003}. Most of the atmospheric X-ray emission of galaxies is observed below $2.0~\rm keV$, while the more energetic photons provide a constraint on the LMXB contribution.

The X-ray emission from both the hot gas and the LMXBs is absorbed by Galactic gas with solar abundances. For each object, we fix the Galactic absorption column density to the value determined by the Leiden/Argentine/Bonn (LAB) Survey of Galactic HI \citep[Table~\ref{tab:nhtot};][]{LAB}. 

\begin{table}
\centering
	\caption{Total hydrogen column densities taken from \citet{LAB}.}
	\begin{tabular}{lc}
		\hline
		galaxy	  &  $N_{\rm H}~ [10^{20}~\rm cm^{-2}]$ \\
		\hline
		NGC 1961 &	11.7	\\
		NGC 3607 & 	1.36	\\
		NGC 3665 & 	2.00	\\
		NGC 4382 & 	2.54	\\
		NGC 4459 & 	2.67	\\
		NGC 4526 &  1.47	\\
		NGC 5353 &	0.954	\\
		\hline
	\end{tabular}
	\label{tab:nhtot}
\end{table}

In the following sections, the results are given with $1\sigma$ error bars.

\section{Results}\label{sec:results}
\subsection{Imaging analysis}\label{sec:imaging}

\begin{figure}
	\centering
	\begin{subfigure}{.32\linewidth}
		\includegraphics[width=\linewidth]{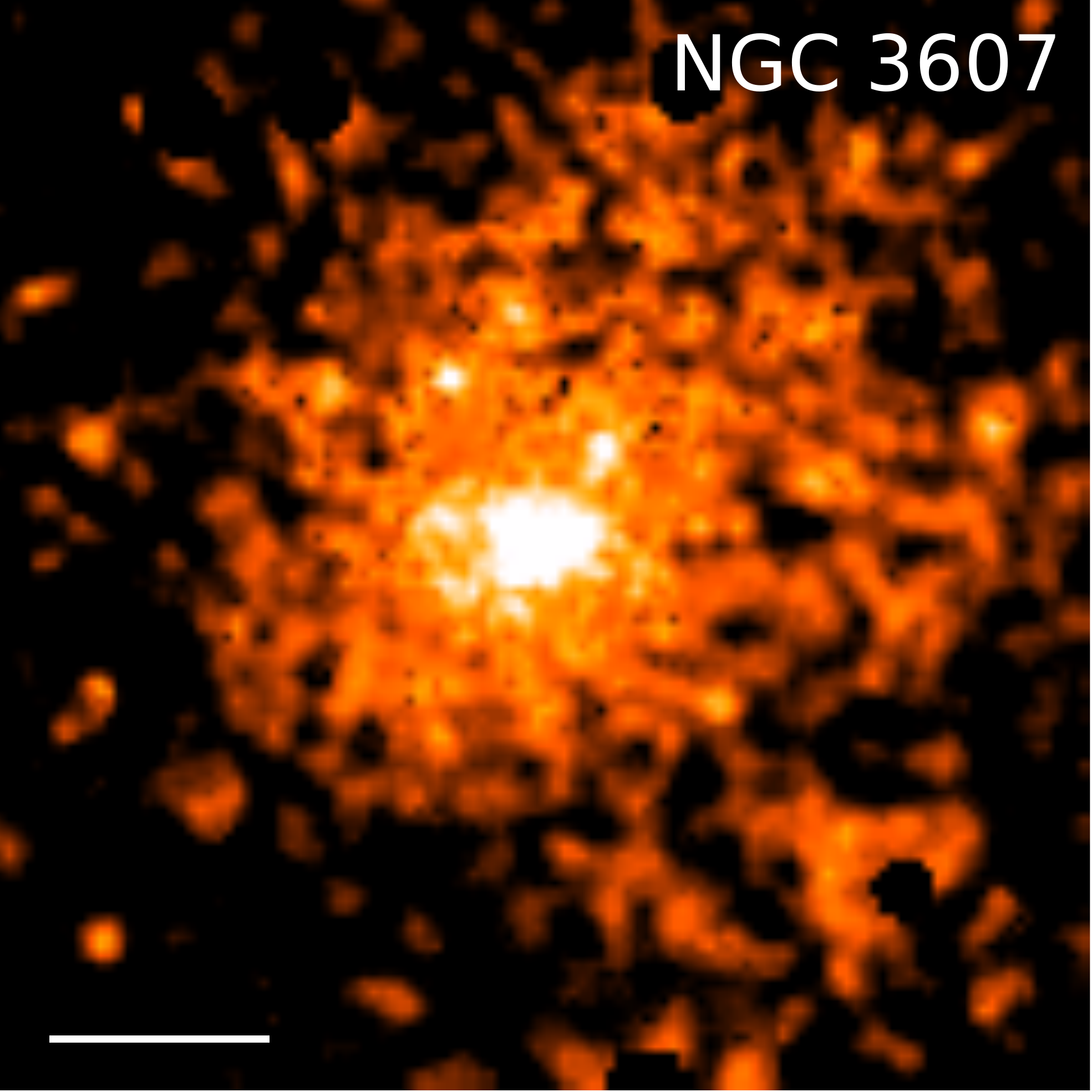}
	\end{subfigure}
	\begin{subfigure}{.32\linewidth}
		\includegraphics[width=\linewidth]{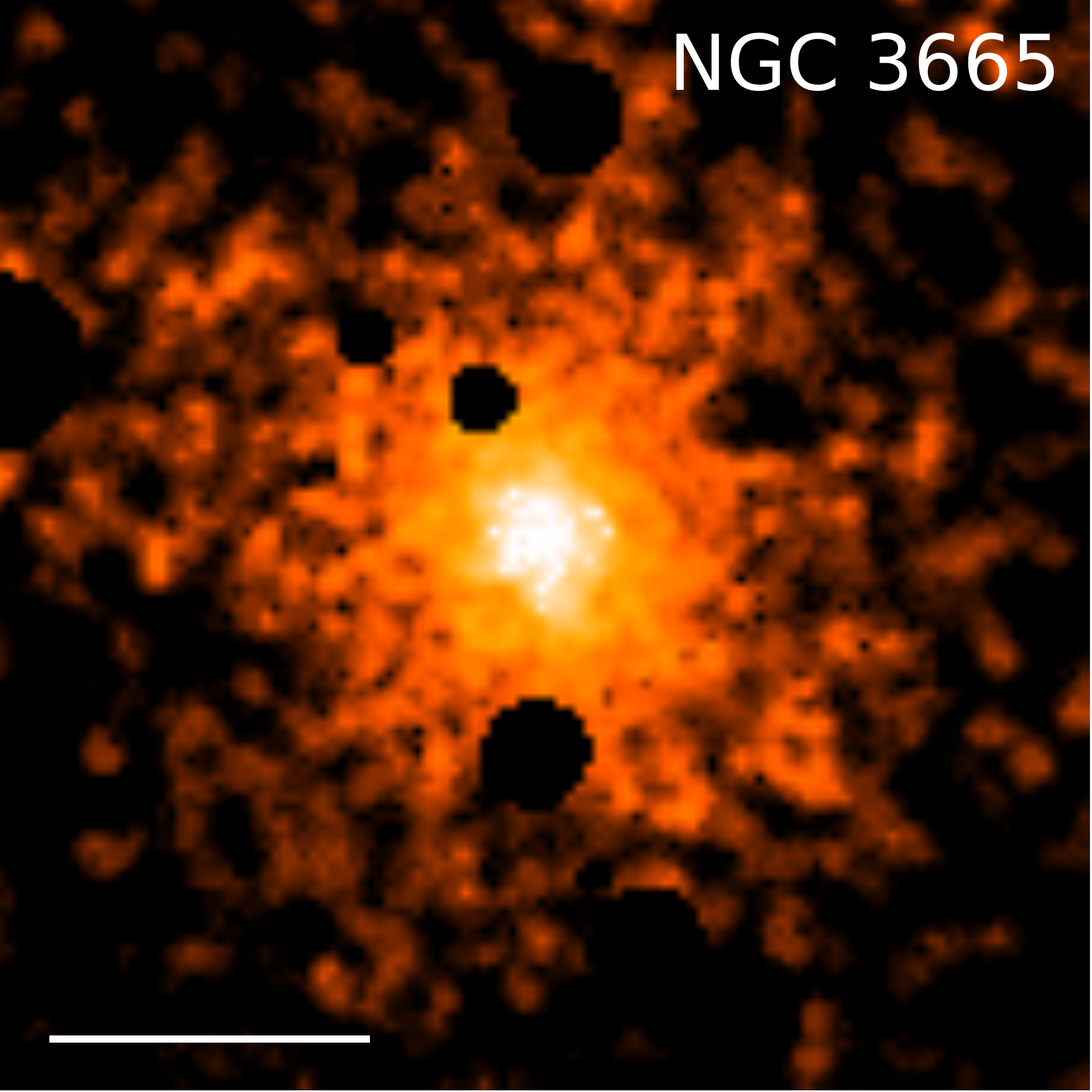}
	\end{subfigure}\vspace{0.05cm}
	\begin{subfigure}{.32\linewidth}
		\includegraphics[width=\linewidth]{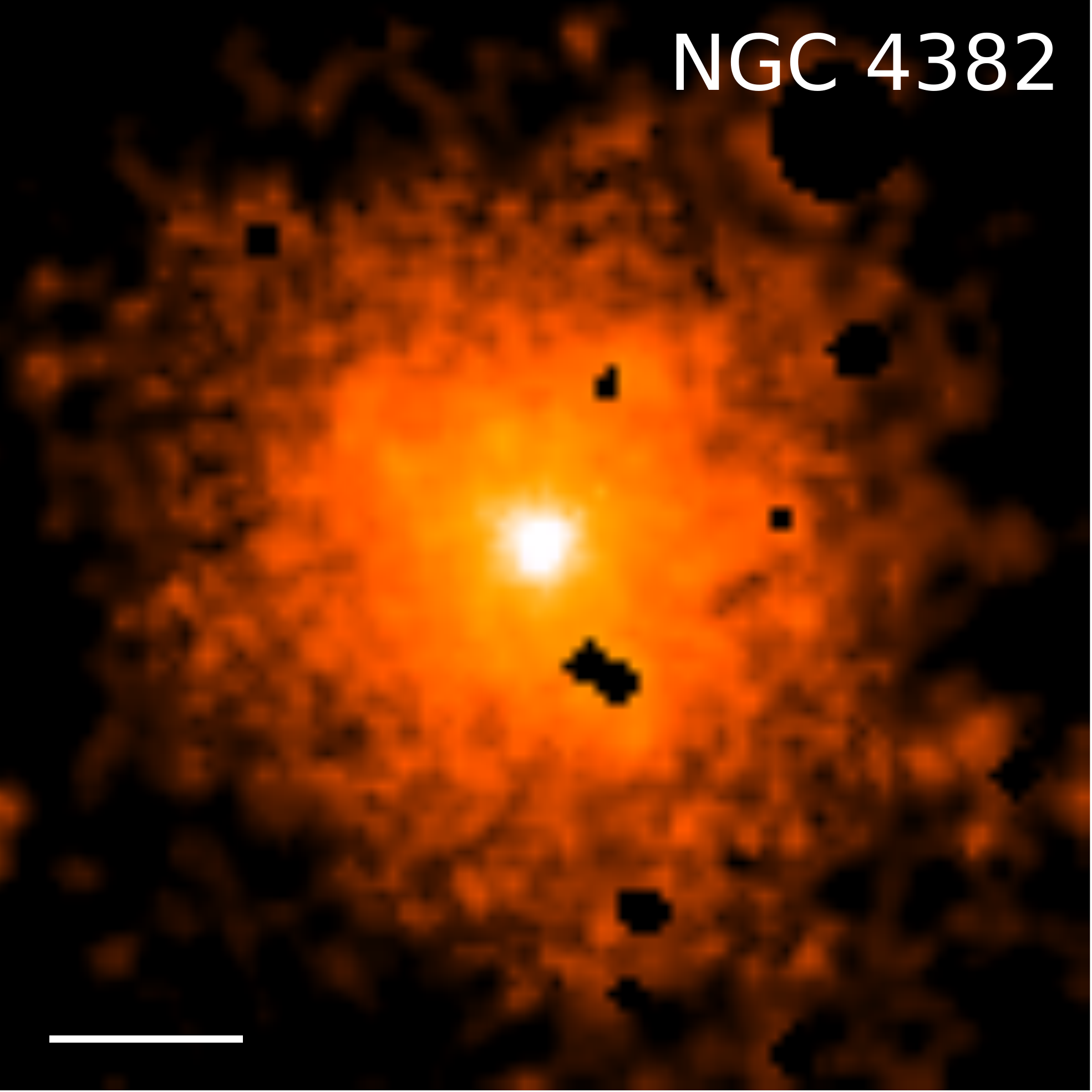}
	\end{subfigure}
	\begin{subfigure}{.32\linewidth}
		\includegraphics[width=\linewidth]{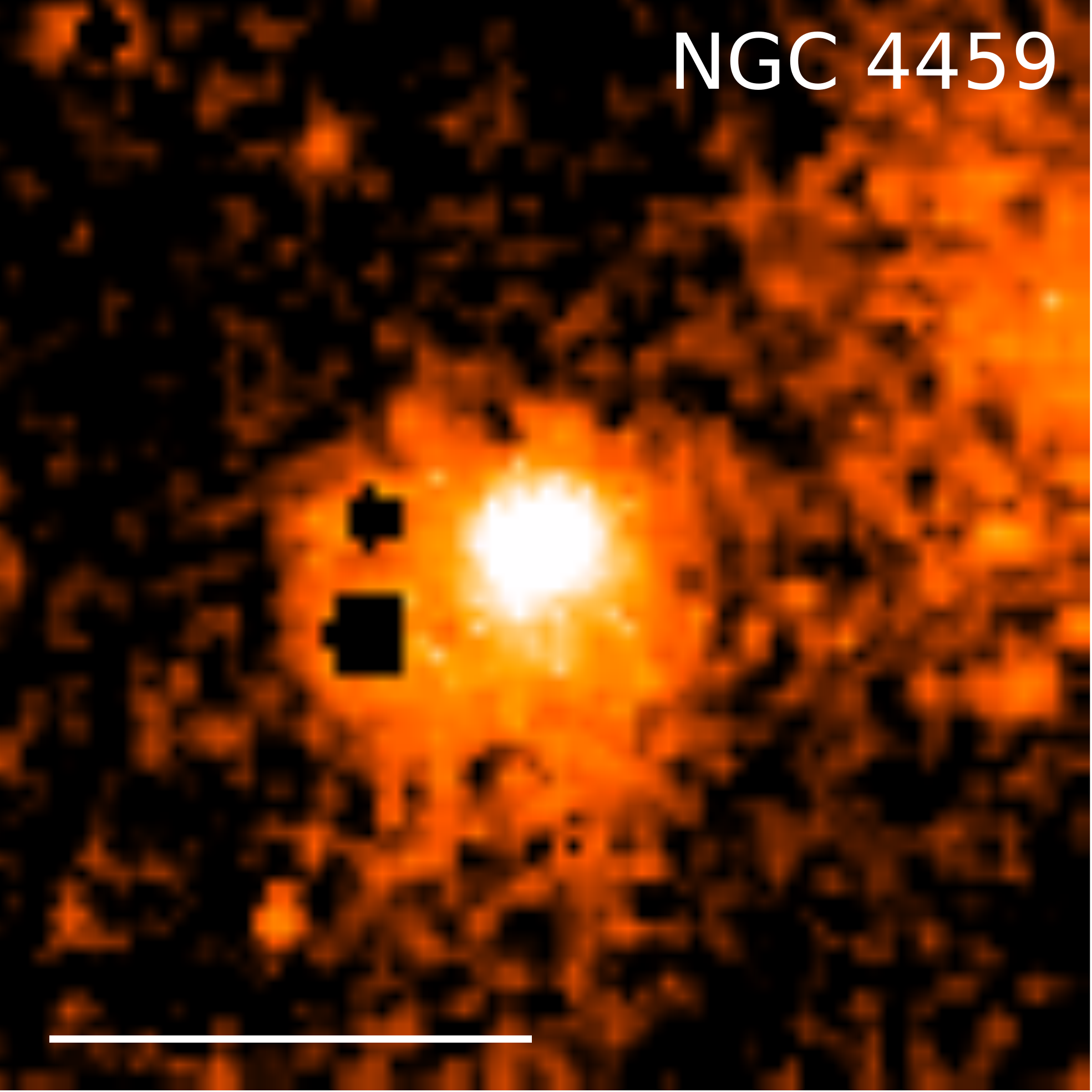}
	\end{subfigure}
	\begin{subfigure}{.32\linewidth}
		\includegraphics[width=\linewidth]{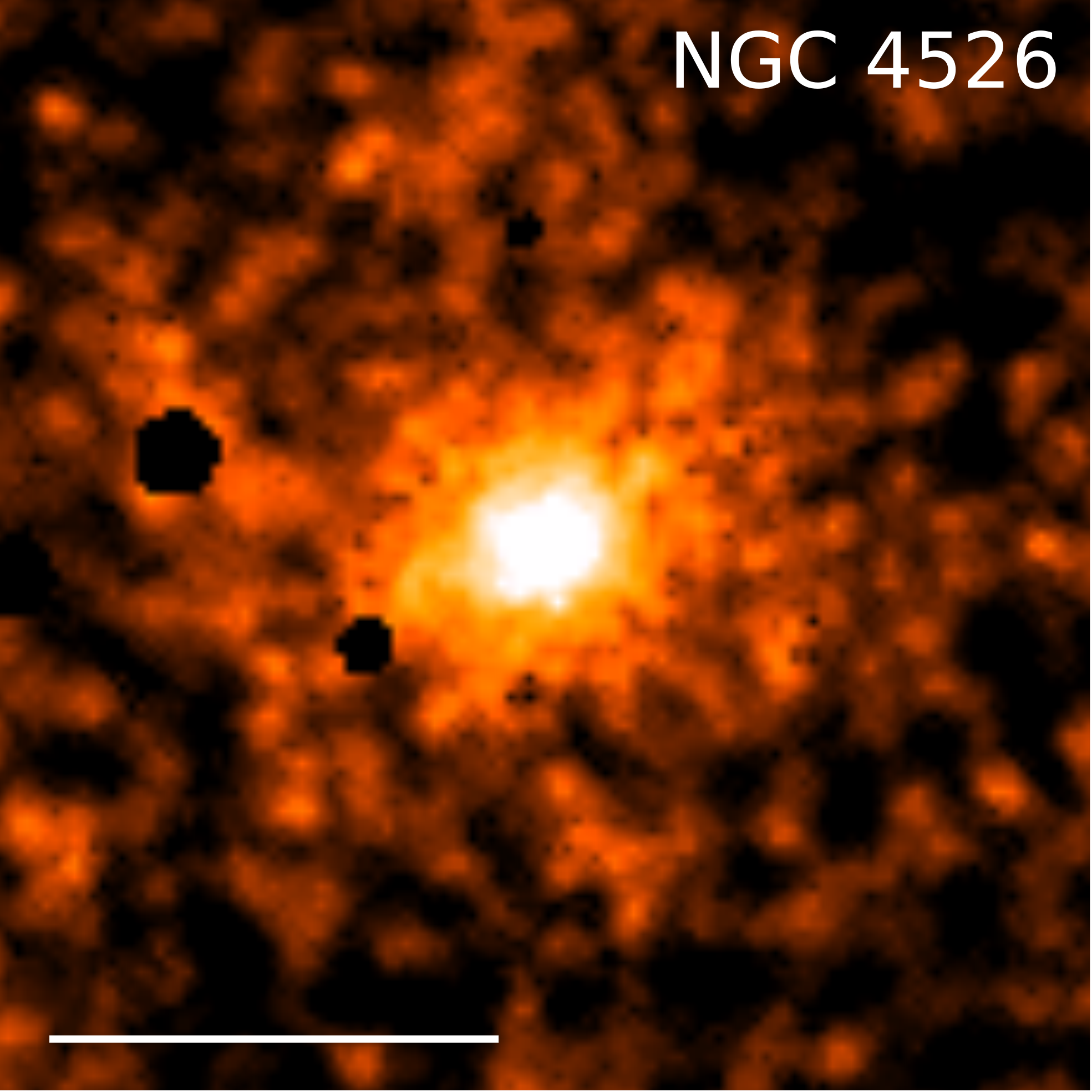}
	\end{subfigure}\vspace{0.05cm}
	\begin{subfigure}{.32\linewidth}
		\includegraphics[width=\linewidth]{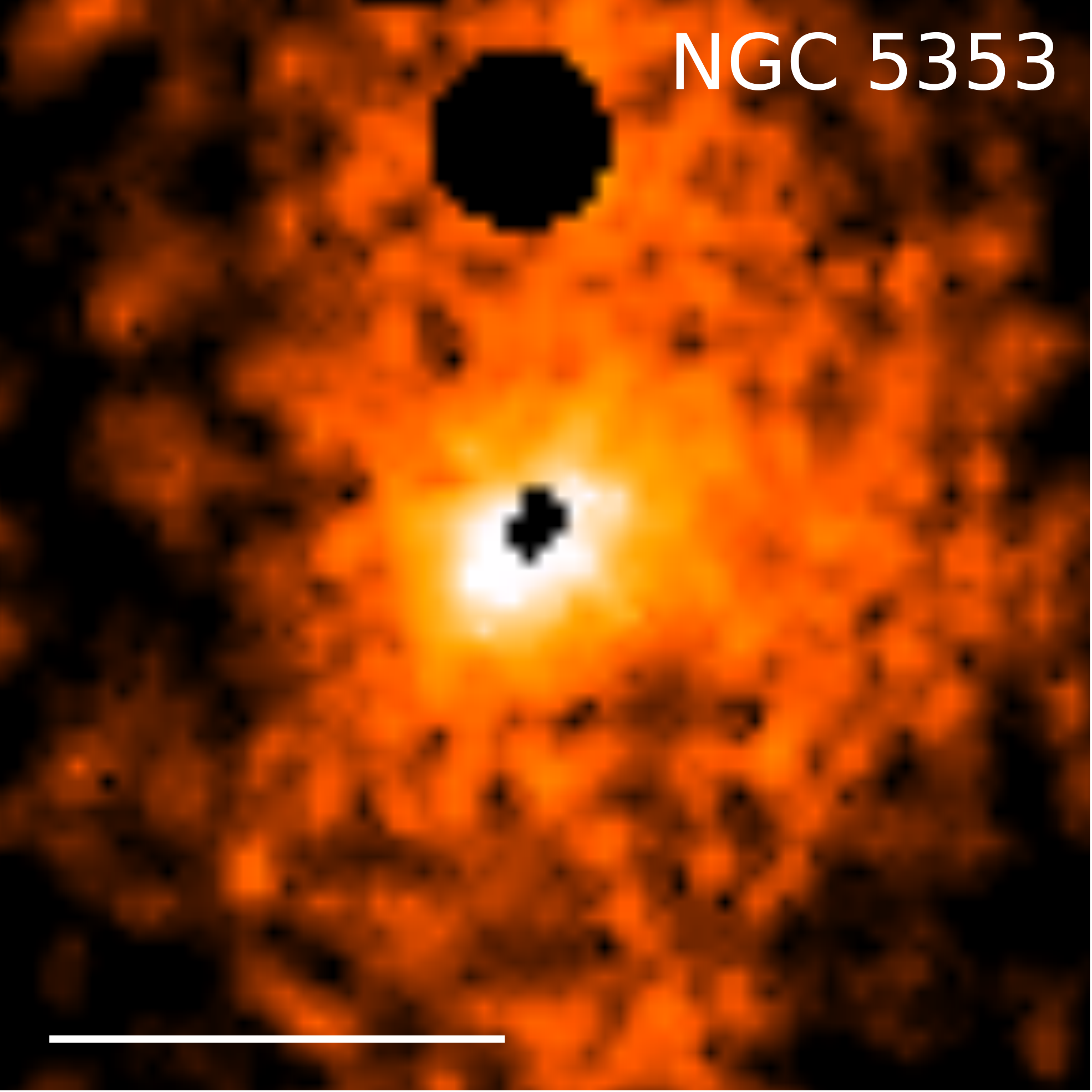}
	\end{subfigure}
	\begin{subfigure}{.32\linewidth}
		\includegraphics[width=\linewidth]{empty-cut.png}
	\end{subfigure}
	\begin{subfigure}{.32\linewidth}
		\includegraphics[width=\linewidth]{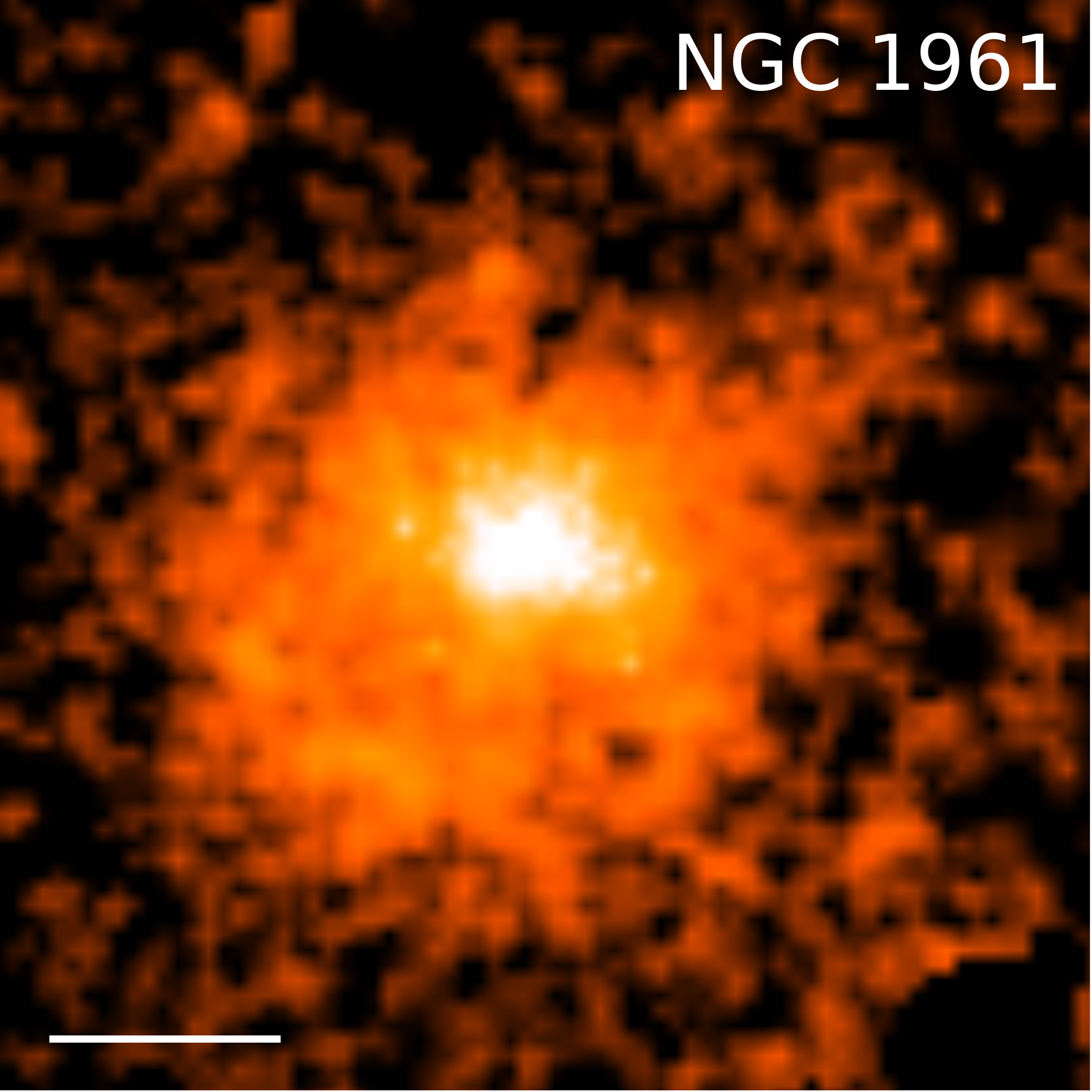}
	\end{subfigure}
	\begin{subfigure}{.32\linewidth}
		\includegraphics[width=\linewidth]{empty-cut.png}
	\end{subfigure}
	\caption{Images of X-ray atmospheres of S0 galaxies in our sample and NGC~1961, all extracted in the energy range $ 0.3-2.0~\rm keV $. The images are displayed on log-scale in order to visualize the full extent of the hot atmospheres, while the most prominent point sources have been removed. The solid line in the lower-left corner of every image represents a scale of 10 kpc.}
	\label{fig:X-ray}
\end{figure}

To study the overall morphology of the galactic atmospheres, we created images in the $ 0.3-2.0~\rm keV $ band. The background subtracted, exposure corrected and adaptively smoothed images are shown in Fig.~\ref{fig:X-ray}. The signal is displayed on a logarithmic scale so that the hot atmospheres are visible out to their outskirts. Bright point sources are removed in the same manner as for the spectral analysis

To determine the flattening of the atmospheres, we used the CIAO \citep[version 4.12, ][]{CIAO} fitting tool Sherpa \citep{Sherpa} to fit each galaxy with a 2D $ \beta $-model \citep{Cavaliere1976, Cavaliere1978} of the form

\begin{equation}\label{eq:beta}
I(r) = I_{\rm c} \left[ 1+ r^2\right]^{-3\beta/2},
\end{equation}

\noindent where

\begin{equation}\label{eq:xybeta}
r^2 = \displaystyle \frac{(1-\varepsilon_{\rm X})^2\tilde{x}^2 + \tilde{y}^2}{r_0^2 (1-\varepsilon_{\rm X})^2}
\end{equation}

\noindent and

\begin{equation}
\begin{aligned}
\tilde{x}&=(x-x_{\rm c})\cos\theta + (y-y_{\rm c})\sin\theta,\\
\tilde{y}&=(y-y_{\rm c})\cos\theta - (x-x_{\rm c})\sin\theta.
\end{aligned}
\label{eq:xy}
\end{equation}

\noindent The normalisation $I_{\rm c}$, the $\beta$ parameter, the centre of the emission $ [x_{\rm c}, y_{\rm c}] $, position angle $ \theta$, core radius $r_0$, and ellipticity $\varepsilon_{\rm X}$ were all left free. The X-ray emission was fitted out to the last annulus presented in this work. Best-fitting values of the $\beta$ parameter, $r_0$, $\varepsilon_{\rm X}$ and position angle $\rm PA_{\rm X} = \theta +{90^\circ}$ are listed in Table~\ref{tab:ellipticity} along with the ellipticity and position angle of the stellar component taken from the literature. It is immediately obvious that the position angles of the X-ray gas and stellar distributions are generally similar. The measured ellipticity is on average lower for the X-ray gas (the mean ratio of the X-ray gas and stellar ellipticity is $\overline{\varepsilon_{\rm X}/\varepsilon_{\star}} = 0.6 \pm 0.3$) and does not correlate directly with the ellipticity of the stellar component.

\begin{table*}
	\caption{X-ray ellipticity ($\varepsilon_{\rm X}$), position angle ($\rm PA_{\rm X}$), parameter $\beta$ and core radius ($r_0$) determined from $\beta$-model fitting and their optical counterparts, $ \varepsilon_{\star} $ and $\rm PA_{\star}$, from \citet{Atlas3DII} and \citet{Jarrett2003} in the case of NGC~1961, for which the uncertainty has not been published.}
	\begin{tabular}{ccccccc}
		\hline
		NGC& $\varepsilon_{\rm X}$ & $\varepsilon_{\star}$ & $\mathrm{PA_X}$ [deg] & $\mathrm{PA}_{\star}$ [deg] & $\beta$ & $r_0$ [pc] \\
		\hline
		3607 & $0.144\pm0.011$ & $0.13\pm0.08$ & $119.7\pm2.3$ & $ 124.8\pm7.6 $ & $0.440\pm0.007$ & $35\pm2$ \\
		3665 & $0.175\pm0.005$ & $0.22\pm0.01$ & $29.0\pm1.0$ & $ 30.9\pm2.0  $ & $0.643\pm0.008$ & $142\pm3$ \\
		4382 & $0.110\pm0.006$ & $0.25\pm0.07$ & $29.8\pm1.7$ & $ 12.3\pm11.0 $ & $ 0.349\pm0.004$ & $31\pm2$ \\
		4459 & $0.060\pm0.016$ & $0.21\pm0.03$ & $134.1\pm7.8$ & $ 105.3\pm1.9 $ & $ 0.79\pm0.06$ & $58\pm4$ \\
		4526 & $0.218\pm0.005$ & $0.76\pm0.05$ & $116.8\pm0.7$ & $ 113.7\pm1.2 $ & $ 1.00\pm0.02$ & $129\pm3$ \\
		5353 & $0.253\pm0.004$ & $0.48\pm0.04$ & $136.6\pm0.6$ & $ 140.4\pm4.9 $ & $ 0.86\pm0.01$ & $157\pm3$ \\
		\hline
		1961 & $0.161\pm0.009$ & 0.330 & $100.8\pm1.8$ & $ 92.0\pm2.0 $ & $ 0.46\pm0.01$ & $ 63\pm3 $ \\
		\hline
	\end{tabular}
	\label{tab:ellipticity}
\end{table*}

The X-ray morphologies of NGC~3607 and NGC~4459 show the presence of a tail indicating ram-pressure stripping by the ambient intra-cluster/group medium or an interaction with a nearby galaxy. An X-ray source seen to the right of NGC~4459 in Fig.~\ref{fig:X-ray} (unrelated emission of a distant galaxy cluster) has been excluded from the analysis.
Regarding past AGN activity, no prominent features are visible in the diffuse emission, nor can be identified in available \textit{Chandra} observations.

\subsection{Thermodynamic properties}\label{sec:thermodynamics}
\subsubsection{Gas mass and temperature}

First, we fitted the X-ray emission within $2 - 6~R_{\rm e}$ for all galaxies with a single- or multi-temperature model, depending on the spectrum (see Sect. \ref{sec:analysis}), to obtain the global properties of the hot gas. The radial extent of the analysed emission was set by the data quality and the maximal radius corresponds to the outer edge of the last annulus presented in this work (see Table \ref{tab:global}). Best-fitting parameters derived from this fit are listed in Table~\ref{tab:global}, where the gaussian width of the temperature distribution of the multi-temperature model is denoted as $ \sigma_{\mathrm{T_{X}}}$ and is presented for spectra where this model provides an improved fit.

We estimated the total gas mass $M_{\rm X}$ of the hot atmosphere by summing the gas masses calculated for each shell from the particle number densities $ n $ derived from the deprojected spectra.

\noindent The summation has been performed out to the last annulus presented in this work and the resulting masses are given in Table~\ref{tab:global}. For the density profiles, see Fig.~\ref{fig:n-single} in the Appendix.

\begin{figure}
	\centering
	\includegraphics[width=0.981\linewidth]{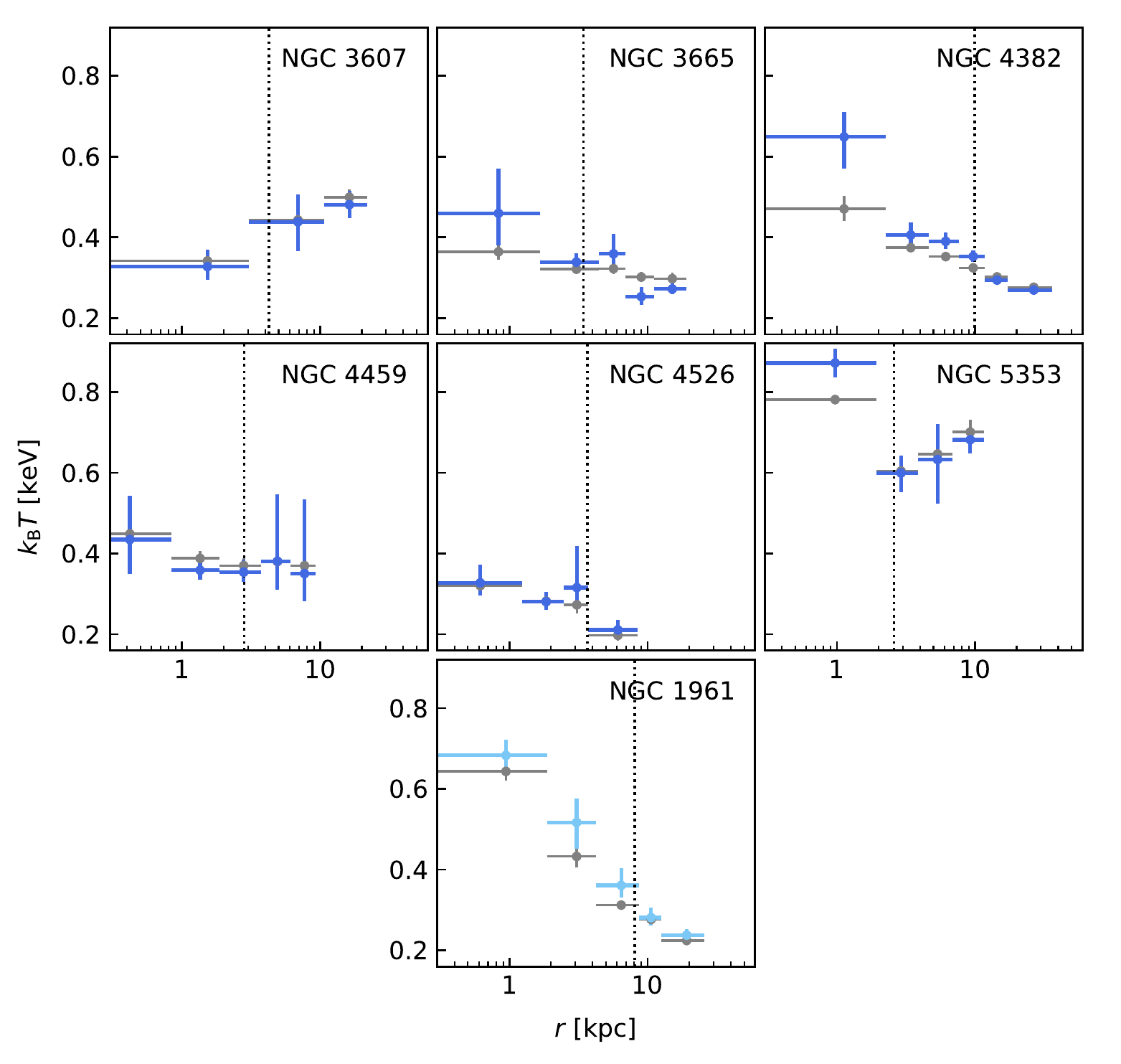}
	\caption{Radial, azimuthally averaged temperature profiles derived from projected (grey) and deprojected (blue) spectra with metallicity fixed at $0.5~\rm Z_{\odot}$. For clarity, dark blue points represent S0 galaxies in our sample, while the profile of the spiral galaxy NGC~1961 is plotted in light blue. The effective radius (see Table \ref{tab:geometry}) is represented by the black dotted line in each panel.}
	\label{fig:kT-single_kpc}
\end{figure}

The best-fitting radial temperature distributions obtained from both projected and deprojected spectra are presented in Fig.~\ref{fig:kT-single_kpc}. 
The spiral galaxy is plotted in a lighter shade of blue to be easily distinguishable from lenticulars.
Monotonically radially decreasing temperature profiles are present in NGC~4382 and NGC~1961, while clearly outwardly increasing temperature is only observed in NGC~3607. The profiles of NGC~3665, 4459 and 4526 do not show any significant trends and within $3\sigma$ uncertainties are close to being isothermal.

To test whether the results are affected by the instrumental PSF, we performed an equivalent analysis using \textit{Chandra} data of NGC~4526 \citep[following the analysis procedures of][]{Lakhchaura2018}, the object with the steepest X-ray brightness profile in our sample. The widths of the analysed annuli are at least 30 times larger than the angular resolution of {\it Chandra}. The obtained results agree with those presented in this work and therefore we expect the \textit{XMM-Newton} PSF to have a negligible effect on our findings.

\begin{table}
	\caption{Total X-ray gas luminosities, emission-weighted temperatures, and gas masses. Where a single temperature model did not provide a good fit, a multi-temperature model with a gaussian emission measure distribution was used and the best-fitting value of the parameter $ \sigma_{T_{\mathrm{X}}}$ is presented. In the last column, the radius within which the spectra were extracted is given.}
	\begin{tabular}{ccccccc}
		\hline
		object & $ L_{\mathrm{X}}$ & $ k_{\mathrm{B}}T_{\mathrm{X}}$ & $ \sigma_{T_{\mathrm{X}}}$ & $ M_{\mathrm{X}}$ \vspace{0.1cm} & $R_{\rm max}$\\
		NGC & $10^{40}~\rm erg~s^{-1}$ & $ \rm keV $ & $ \rm keV $ & $ 10^9~\rm M_{\odot}$ & $\rm kpc$\\
		\hline
		3607 & 1.74 & $ 0.411 ^{+0.025 }_{-0.009 }$ & $-$ & $ 1.52 ^{+0.07 }_{-0.17 }$ & 21.8 \vspace{0.2cm} \\ 
		3665 & 2.30 & $ 0.312 ^{+0.006 }_{-0.006 }$ & $-$ & $ 1.09 ^{+0.07 }_{-0.07 }$ & 19.1 \vspace{0.2cm} \\
		4382 & 7.97 & $ 0.316 ^{+0.025 }_{-0.024 }$ & $ 0.012 ^{+0.017 }_{-0.012 }$ & $ 5.26 ^{+0.10 }_{-0.10 }$ & 36.1 \vspace{0.2cm} \\
		4459 & 0.31 & $ 0.390 ^{+0.041 }_{-0.014 }$ & $-$ & $ 0.12 ^{+0.02 }_{-0.02 }$ & 9.3 \vspace{0.2cm} \\
		4526 & 0.71 & $ 0.260 ^{+0.013 }_{-0.019 }$ & $-$ & $ 0.15 ^{+0.03 }_{-0.02 }$ & 8.5 \vspace{0.2cm} \\
		5353 & 4.21 & $ 0.651 ^{+0.020 }_{-0.020 }$ & $ 0.225 ^{+0.042 }_{-0.046 }$ & $ 0.49 ^{+0.03 }_{-0.03 }$ & 11.7 \\
		\hline
		1961 & 4.79 & $ 0.298 ^{+0.030 }_{ -0.082 }$ & $ 0.202 ^{+0.082 }_{-0.053 }$ & $ 6.01 ^{+0.49 }_{-0.46 }$ & 25.8\\
		\hline
	\end{tabular}
	\label{tab:global}
\end{table}

\subsubsection{Entropy}

A key physical quantity describing the thermodynamic states of hot galactic atmospheres is entropy. We adopt here its definition customary in this field and refer to the entropy index $K$ defined below as the entropy hereafter,

\begin{equation}\label{eq:K}
K \equiv k_{\mathrm{B}}Tn_{\mathrm{e}}^{-2/3}.
\end{equation}

\begin{figure}
	\centering
	\includegraphics[width=1.\linewidth]{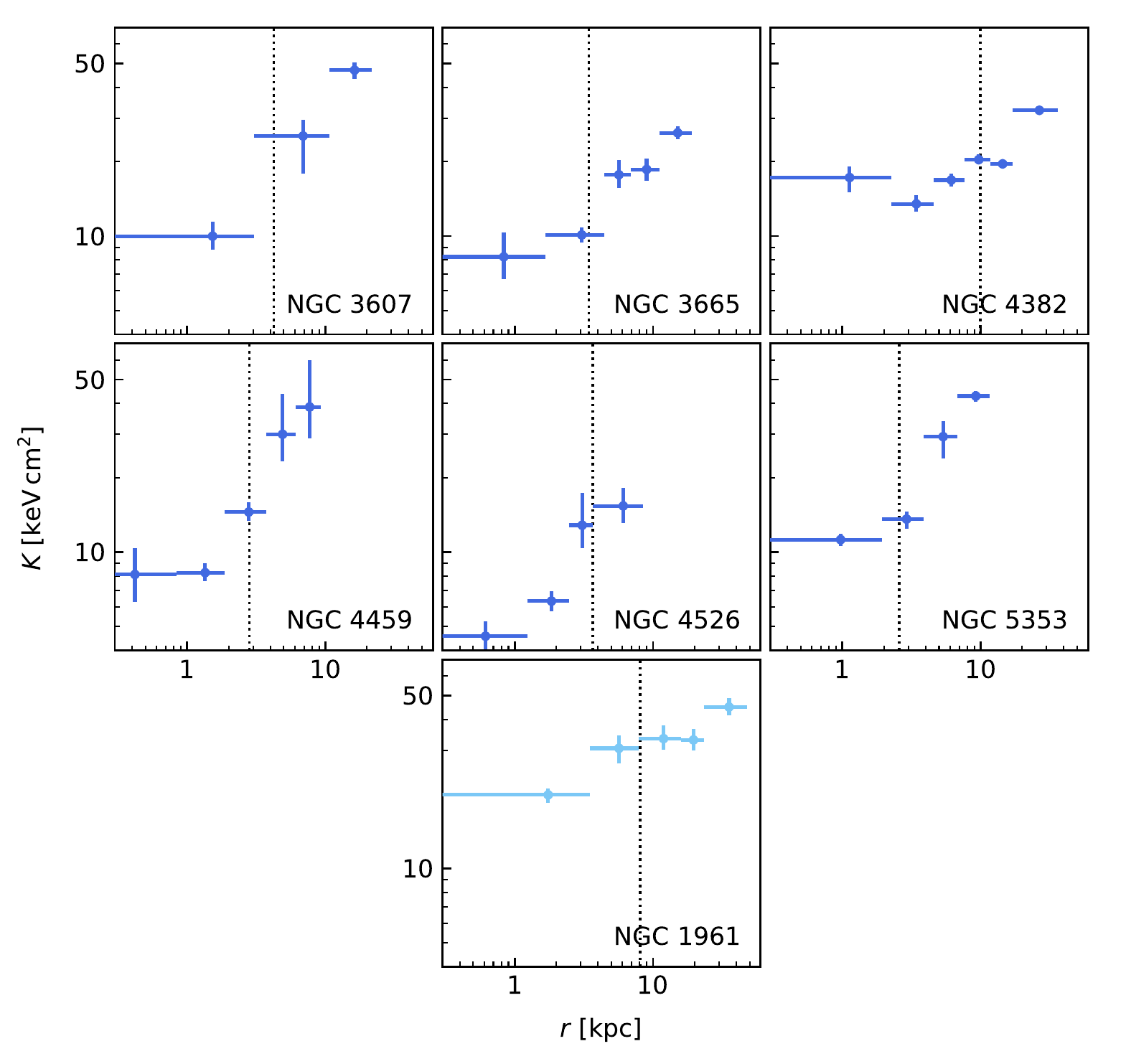}
	\caption{Entropy profiles (see equation \ref{eq:K}) derived from deprojected spectra.}
	\label{fig:K-single}
\end{figure}

\noindent This definition relates to the thermodynamic entropy per particle, $s$, of non-interacting monoatomic particles as $\Delta s = 3/2 k_{\rm B}\ln K$. A gravitationally stratified atmosphere in hydrostatic equilibrium should have an entropy profile rising monotonically with radius, while a flat or decreasing trend would indicate a convectively unstable environment. 

\begin{figure}
	\centering
	\includegraphics[width=0.95\linewidth]{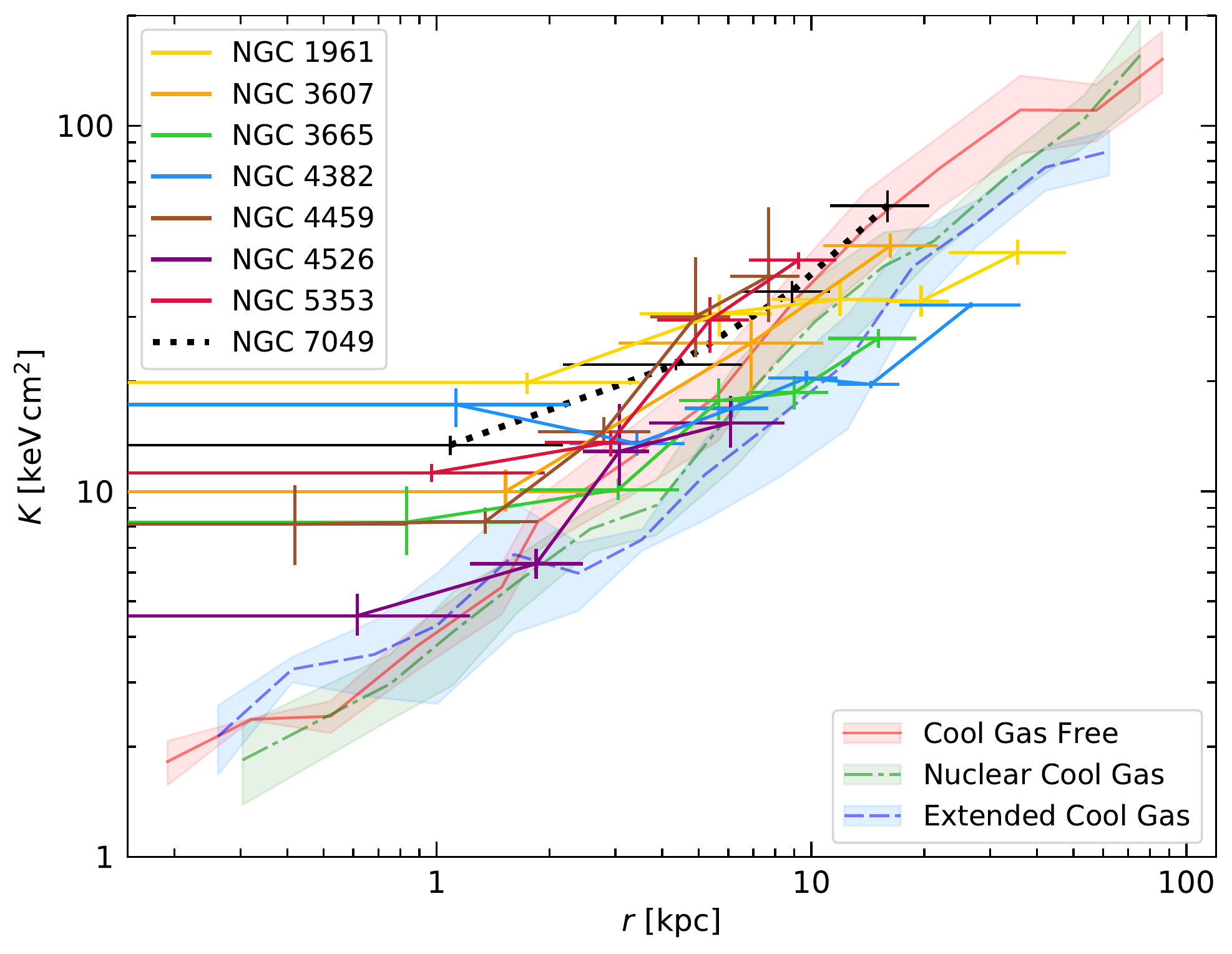}
	\caption{Entropy profiles of the rotationally supported galaxies in this study (solid lines), NGC~7049, an S0 from previous work \citep[black dotted line]{Juranova2019} and a sample elliptical galaxies distinguished by the extent of cool gas of \citet{Lakhchaura2018}. For the ellipticals, lines signify median profiles and surrounding shaded regions the median absolute deviations.}
	\label{fig:K-S0s+Kiran}
\end{figure}

The entropy distribution for all analysed galaxies is plotted in Fig.~\ref{fig:K-single}, separately. For a clearer comparison with elliptical galaxies, the entropy profiles of S0s are plotted in Fig.~\ref{fig:K-S0s+Kiran} together with the results obtained for a sample of 49 giant ellipticals presented by \citet{Lakhchaura2018}.
In general, the central entropy of the lenticulars lies above the values observed in most of the slow-rotating elliptical galaxies.

\begin{figure}
	\centering
	\includegraphics[width=0.95\linewidth]{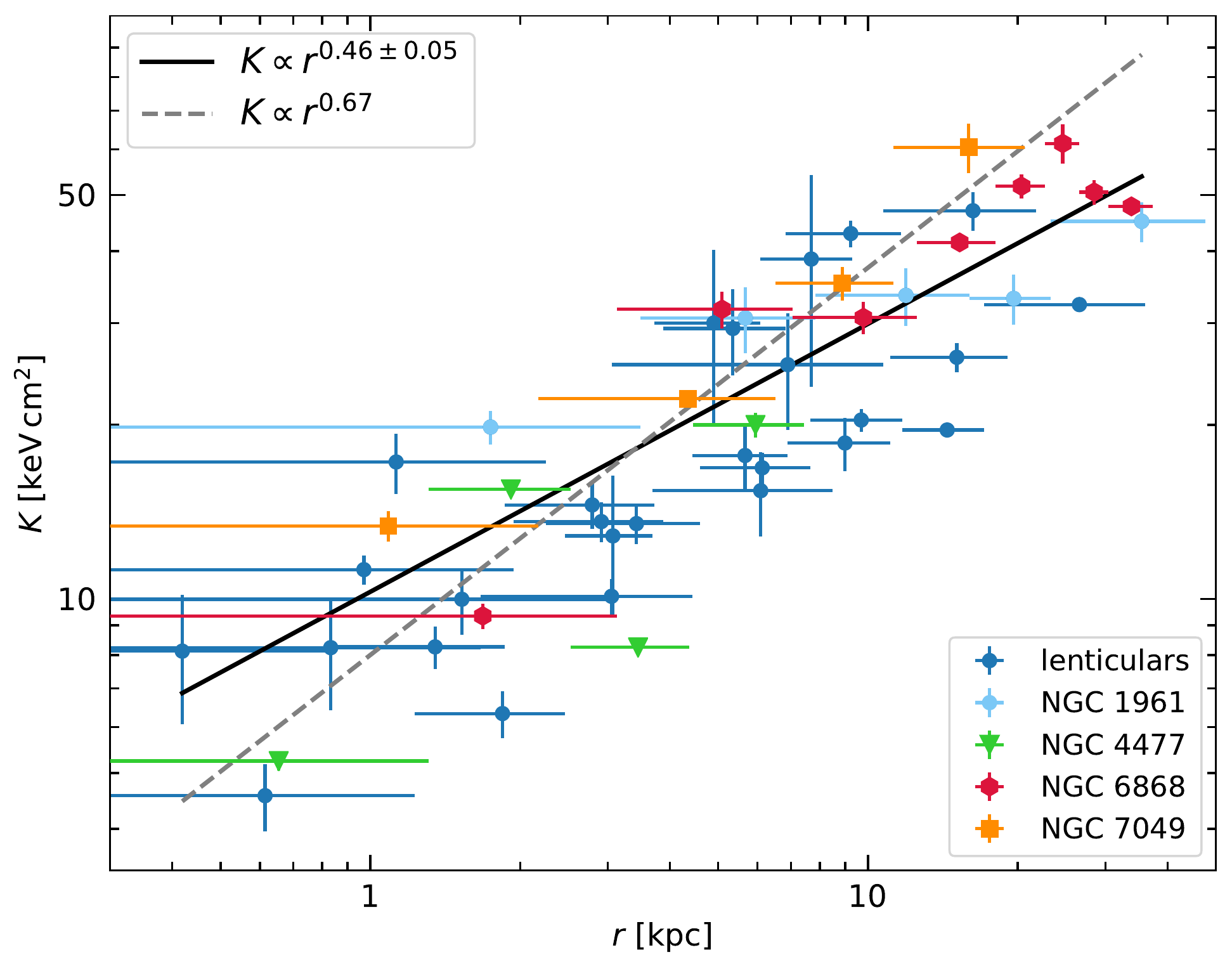}
	\caption{Entropy profiles of rotationally supported galaxies in this study together with profiles of NGC~4477 \citep{Li2018}, NGC~7049 \citep{Juranova2019}, and NGC~6868 \citep{Werner2014, Lakhchaura2018}. The black line represents a best-fitting power law model determined from a fit of all plotted data points. The grey dashed line illustrates the profile usually observed in X-ray atmospheres (see the text for more detailed discussion).}
	\label{fig:K-rot-fit}
\end{figure}

Purely gravitational heating would result in an entropy profile given by $K \propto r^{1.1}$, which is usually not observed due to the heating by the AGN and supernovae that centrally increase the gas entropy, flattening the profiles to $K \propto r^{0.67}$ \citep{Panagoulia2014a, Babyk2018b}. To quantify the amount of flattening in rotating atmospheres, we fitted the average entropy profiles of all S0 galaxies in our sample with a power-law model, together with the fast-rotating spiral NGC~1961 and also including the galaxies NGC~4477 \citep{Li2018}, NGC~7049 \citep{Juranova2019}, and NGC~6868 \citep{Werner2014, Lakhchaura2018}, which show discs of warm/cold gas indicative of rotational support. 
The resulting average power-law index $ \Gamma = 0.46\pm0.05 $ and the corresponding best fitting curve is plotted together with the data in Fig.~\ref{fig:K-rot-fit}. The scatter is, however, large and the flattest entropy profiles are observed in NGC 1961 and NGC 4382, with slopes of $\Gamma=0.24\pm0.05$ and $\Gamma=0.3\pm0.1$, respectively. After excluding these two galaxies and the last two data points of NGC 6868 the average entropy profile becomes $ \Gamma = 0.64\pm0.06 $, consistent with \citet{Babyk2018b}.

\subsection{Thermal stability}\label{sec:stability}
Higher central entropy suggests recent heating and gas expulsion. AGN activity has been confirmed in NGC~3665 and \textit{Chandra} observations show an X-ray point source in the centre of NGC~3607, NGC~3665, NGC~4459, and NGC~4526. Therefore, to address the thermal stability of the gas, we computed cooling time profiles, which are shown in Fig.~\ref{fig:tc-single} in the Appendix. We computed $t_{\rm cool} $ as

\begin{equation}\label{eq:tcool}
t_{\mathrm{cool}} = \frac{3}{2}\frac{(n_{\mathrm{e}} + n_{\mathrm{i}})k_{\mathrm{B}}T}{n_{\mathrm{e}} n_{\mathrm{i}} \mathit{\Lambda}(T, Z)},
\end{equation} 

\noindent where the electron and ion densities are denoted as $ n_{\rm e}$ and $ n_{\rm i}$, respectively, and $\mathit{\Lambda}(T)$ stands for the cooling function. In addition to cooling time alone, profiles of cooling time to free-fall time ratio were derived from the observed thermodynamic properties. The free-fall time was computed under the assumption of hydrostatic equilibrium, using gravitational acceleration

\begin{equation}
g = -\frac{1}{\rho} \frac{\mathrm{d}p}{\mathrm{d}r} = -\frac{1}{nm_{\mathrm{H}}\mu} \frac{\mathrm{d}p}{\mathrm{d}r},
\end{equation}

\noindent where $\mu$ is the mean atomic weight, $\mu = 0.62$, and $m_{\rm H}$ is the mass of a hydrogen atom. The particle number density was calculated as $ n = 1.92~n_{\mathrm{e}} $, where the electron number density, $n_{\mathrm{e}}$, was obtained directly from the spectrum normalisation. The values of the pressure ($p=nk_{\rm B}T$) gradient were determined from a fit by a $\beta$-profile or a power-law, when required by the shape of the pressure profile. The resulting values are presented in Fig.~\ref{fig:tctff-sample} for all eight studied objects and separately in the Appendix, Fig.~\ref{fig:tctff-single}. The value of $t_{\rm cool}/t_{\rm ff} = 10 $ is visualised through a dashed grey line and is exceeded at all radii. Despite the confirmed presence of cold gas, the values of $t_{\rm cool}/t_{\rm ff}$ at small radii correspond to values observed in cool gas free galaxies \citep[e.g.,][]{Voit2015b, Hogan2017}.

\begin{figure}
	\centering
	\includegraphics[width=0.95\linewidth]{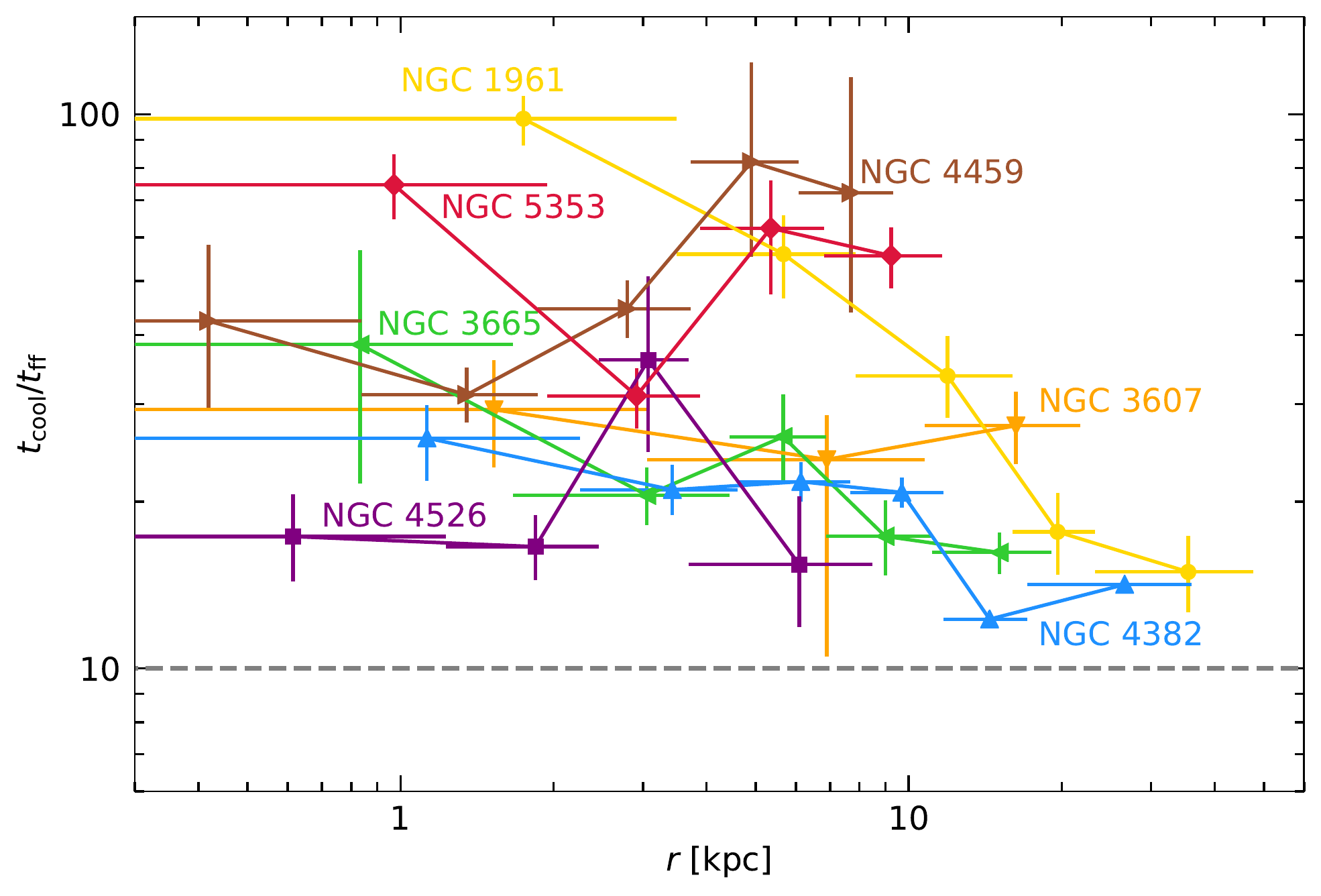}
	\caption{Ratio of cooling time to free-fall time for all studied galaxies. The value $t_{\rm cool}/t_{\rm ff} \approx 10$ (see section \ref{sec:stability}) is visualised as a dashed grey line.}
	\label{fig:tctff-sample}
\end{figure}

\begin{figure}
	\centering
	\includegraphics[width=0.95\linewidth]{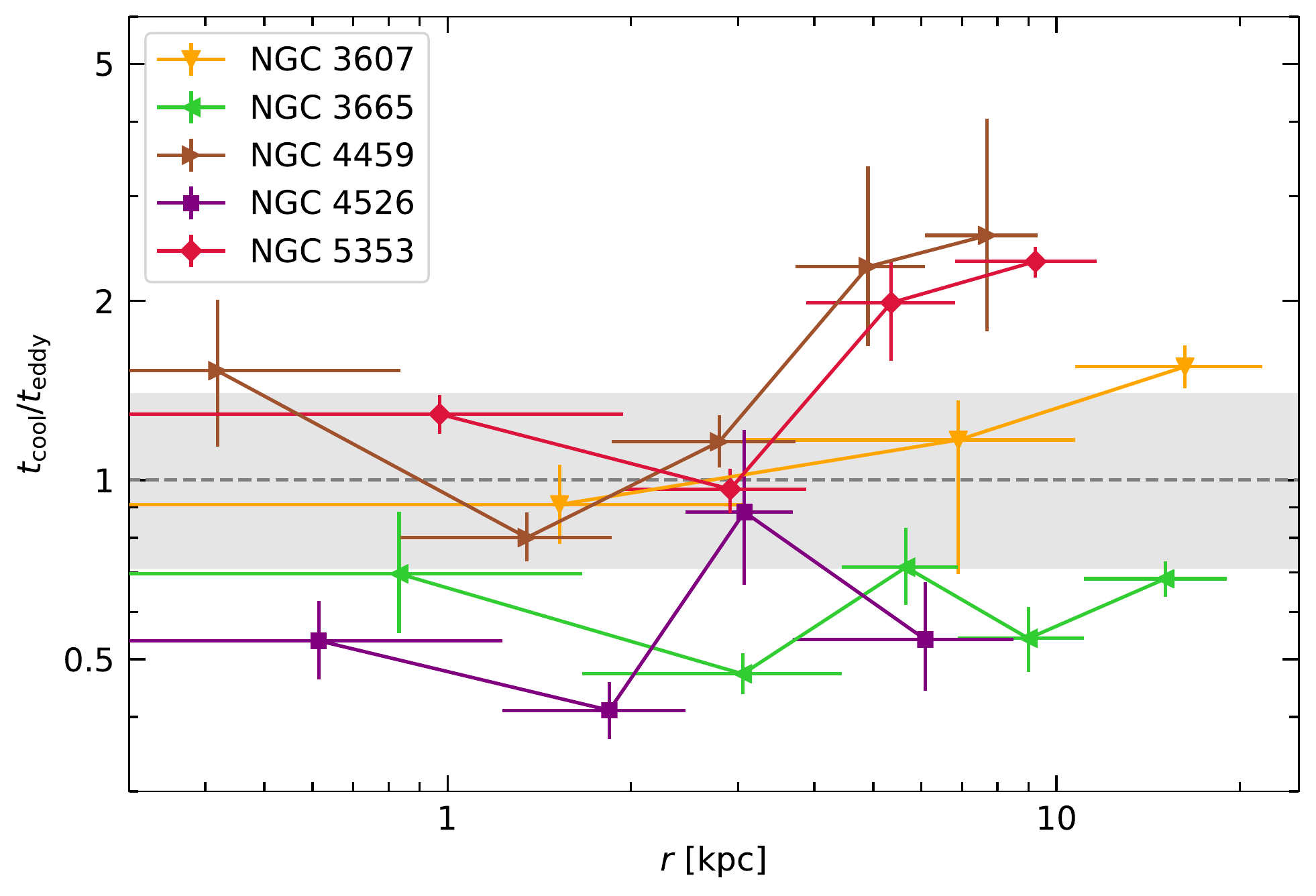}
	\caption{The $C$-ratio (see section \ref{sec:stability}) of S0 galaxies possessing cold gas. The grey region represents the 1$\sigma$ confidence region \citep[from hydrodynamical simulations;][]{Gaspari2018} where  conditions for the development of multiphase condensations are expected to be favourable.}
	\label{fig:C-ratio-sample}
\end{figure}

We also inspected the thermal stability of the hot gas using a parameter that compares the cooling time with a timescale related to turbulent motions in the gas. It has been proposed that when these two timescales become comparable, the physical conditions in the hot atmosphere should favour development of multiphase gas. We compute this quantity, the so-called $C$-ratio, as $ C \equiv {t_{\rm cool}}/{t_{\rm eddy}} $ \citep[see section 5.2. in ][]{Gaspari2018}, where

\begin{equation}
   t_{\rm eddy} = 2\upi \frac{r^{2/3} L^{1/3}}{\sigma_{v,L}}
\end{equation}

\noindent and where $\sigma_{v,L}$ is the gas velocity dispersion at injection scale $L$ (typically 5-10 kpc). According to \citet{Gaspari2018}, this distance can be estimated as a diameter of the cold/warm phase, and the corresponding $\sigma_{v,L} $ can be obtained by extrapolating the measured $\sigma_{v}$ of the cold gas. For the galaxies in our sample, the velocity dispersions in the cold gas have not been published. Nevertheless, to have at least an estimate of this condensation parameter, we adopt the value of the velocity dispersion measured in NGC~7049, $\sigma_{v,L} = 36~\rm km~s^{-1}$ \citep{Juranova2019}. The results computed for S0 galaxies with discs of cold gas are plotted in Fig.~\ref{fig:C-ratio-sample} and show that the conditions in these hot atmospheres are favourable for condensation from the hot phase. We note that the multiphase condensation is expected to occur along helical orbits settling onto a disc where the gas rotational speed exceeds the gas velocity dispersion, i.e. turbulent Taylor number ${\rm Ta_t} > 1$ \citep{Gaspari2015}. Conversely, and typically within the inner kpc region \citep[see NGC 7049;][]{Juranova2019}, we expect ${\rm Ta_t} < 1$ and thus a more chaotic cold accretion rain. Forthcoming high-resolution optical and radio observations of warm and cold gas (e.g. via MUSE and ALMA) will be
key to unveil the detailed top-down multiphase condensation kinematics.

\section{Discussion}\label{sec:discussion}
\subsection{Hot gas morphology and mass}
The projected flattening of the hot atmospheres has been found to be similar to or smaller than that of the stellar component in all studied objects (see Table \ref{tab:ellipticity}), in agreement with the hydrodynamical simulations of e.g., \citet[][]{Brighenti2009, Negri2014b, Gaspari2015}. The ellipticity has been measured using background and point source subtracted X-ray images, which are, however, contaminated by the X-ray emission of unresolved stellar sources, potentially contributing to the observed flattening. However, given that the X-ray emission in the $0.3-2.0~\rm keV$ band is dominated by the gaseous atmosphere, the ellipticity is primarily due to the distribution of the diffuse gas. The principal axes of the ellipsoidal isophotes fitted to the X-ray and optical emission are aligned within the measured uncertainties. This result is consistent with ordered rotation of the X-ray atmosphere in a generally rounder total gravitational potential. We note that, however, a similar effect would also be expected for non-rotating gas in a gravitational potential that is flattened due to rotation.

The assumption of spherical symmetry in the subsequent analysis is not expected to affect our results significantly. The ellipticity of the X-ray atmospheres is low and the logarithmic scale in Fig.~\ref{fig:X-ray} strongly emphasizes features deviating from the azimuthal symmetry of the atmospheres. 

The amount of observed hot gas in S0 galaxies, which has been derived from deprojected densities out to $2 - 6~R_{\rm e}$, varies from $10^8$ to $5\times 10^9~\rm M_{\odot}$. 
Several authors \citep[e.g.,][]{Brighenti2009, Negri2014b, Gaspari2015} have shown that the hot atmospheres of rotationally supported galaxies have smaller hot gas content, lower emission-weighted temperatures and thus lower X-ray luminosities. The largest amount of atmospheric gas has been found in NGC~1961 and NGC~4382, and is comparable to the hot gas content in slow-rotating giant ellipticals \citep{werner2012,Babyk2019}. 

\subsection{Thermodynamic properties}
\subsubsection{Temperature}

Radial azimuthally averaged profiles reveal that the hot gas temperature is systemically lower in the S0 galaxies than in giant ellipticals \citep[see e.g.][]{Lakhchaura2018}. This is an expected outcome, as the virial temperature of less massive S0 galaxies is lower and the ordered stellar motion in rotating systems leads to less effective heating of the gas ejected in stellar mass loss. The radially decreasing temperature profile of NGC~1961 suggests central heating, which can be provided by adiabatic heating, the AGN or by supernovae, exploding relatively frequently due to the star formation of $\sim 10~\rm M_{\odot}~yr^{-1}$. 

A negative temperature gradient is observable also in NGC~4382.
The energy input via type Ia SNe can be estimated from their expected rate, which for a galaxy of this stellar mass \citep[$ M_{\star} = 4\times 10^{11}~\rm M_{\odot}$;][]{Gallo2010} and SFR (see Table~\ref{tab:content}) corresponds to approximately $0.02~\rm yr^{-1}$ using a relation from \citet{Sullivan2006}, or $0.01~\rm yr^{-1}$, when derived from the $B$-band luminosity \citep{Pellegrini2012a}. Assuming a kinetic energy of $E_{\rm SNIa} \approx10^{51}~\rm erg$ per supernova, this rate corresponds to a time-averaged energy injection of $6\times 10^{41}~\rm erg~s^{-1}$ or $3\times 10^{41}~\rm erg~s^{-1}$, respectively. Such heating would be sufficient to compensate for the energy losses of the hot gas if the efficiency of SN heating was at least $\sim 12~\%$ or $\sim 27~\%$, respectively. An additional contribution of the merger event that led to the creation of shells is expected to be of lesser importance. According to cosmological numerical simulations analysed by \citet{Pop2018}, the initial interaction of the progenitors occurred $4-8~\rm Gyr$ ago and most of the stars were stripped $\sim 2~\rm Gyr$ ago, while the central cooling time measured in NGC~4382 is $t_{\rm cool}\approx 0.5~\rm Gyr$. 

\subsubsection{Entropy}

The central entropy in most of the studied S0 galaxies is above that of elliptical galaxies, indicating the presence of a centrally positioned heating mechanism. The low star formation rate suggests that energy injection via winds of young stars and core-collapse supernovae is most likely incapable of providing the required energy. A plausible source of energy could be the central AGN or type Ia supernovae.

The average radial entropy distribution of the rotating galaxies in our sample shown in Fig.~\ref{fig:K-rot-fit} appears to be shallower than the entropy profiles in galaxy clusters and slow-rotating giant ellipticals. The power-law index of the average radial profile is $\Gamma = 0.46\pm0.05 $.
The scatter in the profiles is, however, large and the dominant heating mechanism might differ from galaxy to galaxy.  The flattest entropy profiles are observed in NGC 4382 ($\Gamma = 0.3\pm0.1$) and the massive spiral galaxy NGC 1961 ($\Gamma = 0.24\pm0.05 $). The entropy profiles for the rest of the sample have an index $\Gamma = 0.64\pm0.06$, consistent with the results of \citet{Babyk2018b}. However, the central entropy for all systems in our sample appears higher than the typical values measured in slow rotating giant ellipticals \citep{Lakhchaura2018}.

The relatively flat entropy profiles are consistent with the presence of outflows in the rotationally supported atmospheres. As the outflows redistribute the gas from the central region towards the outskirts, reducing the central gas-mass fraction, the entropy profiles will become shallower. This can be achieved more easily if rotation lowers the effective binding energy (albeit anisotropically) and thus facilitates matter ejection from the core regions.

\subsection{Thermal stability}

The central cooling times do not exceed $\sim 0.5~\rm Gyr$ and remain as low as $\sim 1~\rm Gyr$ out to $ 10~\rm kpc$ in all investigated galaxies, confirming that feedback is necessary to stabilise these hot atmospheres. The observed cooling times at $ 10~\rm kpc$ are consistent with the findings of \citet{Babyk2019}. These results, along with direct traces of AGN activity in several objects in our sample, raise the question of thermal stability of the hot gas.

The calculations of the $C$-ratio \citep{Gaspari2018} shown in Fig. \ref{fig:C-ratio-sample} suggest that turbulence could, in principle, be capable of generating density fluctuations prone to cooling. This conclusion depends on our assumption of the velocity dispersion, as the $t_{\rm eddy}$ scales with $\sigma_{v,L}^{-1}$. Velocity dispersions have been estimated from resonant scattering and line broadening by \citet{Ogorzalek2017} in several elliptical galaxies using spectra from the \textit{XMM-Newton} Reflection Grating Spectrometer. The average best-fitting 3D $\sigma_v$ is approximately $ 190~\rm km~s^{-1}$ \citep[however, none of the galaxies studied by][is a fast rotator]{Ogorzalek2017}.

Rotational support has a radical effect on the thermal stability of the hot gas. In a spherical system, with no rotation, the negative buoyancy of a cool gas clump causes it to fall inward towards the place where it is neutrally buoyant. At that location, the mean heating rate matches the cooling rate, so it is in thermal balance. Even if the heating from a central AGN is anisotropic, the tendency for the gas to find the place where it is neutrally buoyant will cause gas that is overheated to move outward and be replaced by inflowing gas that has previously been underheated, bringing that gas closer to the AGN, where the heating is greater, potentially driving large scale circulation. As a result, it is relatively difficult for thermal instability to grow, whenever $t_{\rm cool}$ significantly exceeds $t_{\rm ff}$ (i.e. TI-ratio $\gg 1$).

This may change when there is appreciable rotational support. A gas clump will generally not have the same density as the ambient gas at the place where it is dynamically stable, so that it no longer naturally finds the place where the ambient heating rate matches its cooling rate. Since the gas distribution is no longer spherical, a spherically symmetric heating rate cannot maintain thermal balance. Rotating gas will tend to cool into a disc, rather than being fed into the AGN. In the absence of significant stellar feedback, this means that there is far more opportunity for the gas to cool into a disc. Indeed, a more relevant condensation criterion in this rotating scenario is the $C$-ratio. As shown in Fig. \ref{fig:C-ratio-sample}, this is found to hover around $C$-ratio $\sim 1$, which is the condensation zone for direct turbulent condensation \citep{Gaspari2018}. This can occur independently of very high TI-ratios (up to 100, as found here), with the formation of a multiphase disc or a chaotic cold accretion rain set by ${\rm Ta_t} > 1$ or $<1$, respectively.

Interestingly, the two galaxies with the greatest gas mass (NGC 1961 and NGC 4382) have the shallowest entropy profiles and the greatest entropy levels in their centres. Assuming that the atmospheres of these galaxies are close to hydrostatic equilibrium, their nearly isentropic profiles will be responsible for the observed pronounced central temperature peaks. 

The shallow entropy profiles are most likely the result of strong central feedback activity. For stellar feedback, where heating is dominated by core collapse supernovae, to operate, stars would need to have formed within the past cooling time. However, as shown in Table \ref{tab:content}, there is no evidence for significant recent star-formation in the investigated galaxies. Once the feedback raises the central entropy level, the heat input from SNIa can exceed the radiative cooling rate within a radius of several kpc \citep{Voit2015b,voit2020} and the energy input of SN Ia from older stars (the SN Ia activity is not part of any feedback loop) might even help to drive galactic winds.

The cold gas reservoir could, in principle, also be replenished by stellar ejecta. According to e.g. \citet{Voit2011}, the gas ejected from stars should heat up and mix with the hot phase in systems with as low central density and star formation rate as in our investigated S0 galaxies. However, the presence of cold gas with PAH molecules suggests a non-negligible role of winds of asymptotic giant branch stars (AGB) in the cold gas production. A plausible explanation has been recently presented by \citet{Li2019} based on 3D hydrodynamical simulations. They suggest that cooling from the hot phase could be induced in the mixing layer of the dusty stellar wind and the surrounding hot gas, leading to the preservation of these fragile particles. 

\section{Conclusions}\label{sec:conclusions}

We presented a study of the X-ray emitting atmospheres of six lenticulars and one massive spiral galaxy to study the effects of angular momentum on the properties of the hot gas. We compare the measured thermodynamic properties with a sample of ellipticals in which the rotational support is negligible. 

We find an alignment between the hot gas and the stellar distribution, with the ellipticity of the X-ray emission generally lower than that of the optical stellar emission, consistent with theoretical predictions for rotationally supported hot atmospheres. 
Two galaxies in the investigated sample, NGC 4382 and the massive spiral NGC 1961 have remarkably flat entropy profiles with $\Gamma\approx 0.2-0.3$, suggesting the presence of outflows in the central regions of these systems. 
These two galaxies also have significantly higher atmospheric gas masses than the other systems in our sample and their gas content is similar to that observed in slow rotating giant ellipticals \citep[e.g.][]{werner2012,Babyk2019}.
The entropy profiles for the rest of the sample have an index $\Gamma = 0.64\pm0.06$, consistent with the results of \citet{Babyk2018b}. However, the central entropy for all systems in our sample appears higher than the typical values measured in slow rotating giant ellipticals \citep{Lakhchaura2018}.

Investigating the relation between hot atmospheres and the observed cold gas phases, we present the derived dimensionless parameters broadly adopted as being related to the thermal stability of hot atmospheres. We obtain a ratio of cooling time to free-fall time of $ t_{\rm cool}/t_{\rm ff} \gtrsim 10$ for all objects. We also estimate the ratio of the cooling and turbulent time-scales and discuss the possibility that the discs of cold gas present in these objects have condensed out from the hot atmospheres.

\section*{Acknowledgements}
N. W. was supported by the Lend\"{u}let LP2016-11 grant awarded by the Hungarian Academy of Sciences and is currently supported by the MUNI Award for Science and Humanities funded by the Grant Agency of Masaryk University. 
M. G. is supported by the Lyman Spitzer Jr. Fellowship (Princeton University) and by NASA Chandra GO8-19104X/GO9-20114X and HST GO-15890.020-A grants. Based on observations obtained with XMM–Newton, an European Space Agency (ESA) science mission with instruments and contributions directly funded by ESA Member States and NASA. This research has made use of the NASA/IPAC Extragalactic Database (NED), which is operated by the Jet Propulsion Laboratory, California Institute of
Technology, under contract with NASA, and software provided by the Chandra X-ray Center in the application packages CIAO and SHERPA.

\section*{Data availability}
The data underlying this article will be shared on reasonable request to the corresponding author.

%%%%%%%%%%%%%%%%%%%%%%%%%%%%%%%%%%%%%%%%%%%%%%%%%%


\bibliographystyle{mnras}
\bibliography{ms}

%%%%%%%%%%%%%%%%%%%%%%%%%%%%%%%%%%%%%%%%%%%%%%%%%%

%%%%%%%%%%%%%%%%% APPENDICES %%%%%%%%%%%%%%%%%%%%%

\appendix

\section{Supplementary material}

\begin{table*}
	\caption{List of observations for each galaxy and exposure times. In the second column, observation ID is given, for which the total observation time $t_{\rm tot}$ in MOS1, MOS2 and pn detectors is shown in columns 3, 4, and 5, respectively. The useful exposure time $t_{\rm net}$ that is not contaminated by soft-proton flaring is given in the next 3 columns. The sum of the flaring-excluded exposure times in all instruments together is given in the last column.}
	\begin{tabular}{cccccccccc}
		\hline
		Object & OBSID & \multicolumn{3}{c}{$t_{\rm tot}~[\rm ks]$} && \multicolumn{3}{c}{$t_{\rm net}~[\rm ks]$} & $\sum t_{\rm net}~[\rm ks]$ \\
		&  & M1 & M2 & pn && M1 & M2 & pn & \\
		\hline
		NGC~1961 	& 0673170101 & 31.5 & 31.5 & 30.4 && 21.5 & 23.5 & 11.3 & \\
		& 0673170301 & 35.0 & 35.1 & 34.0 && 21.2 & 21.1 & 17.5 & \\
		& 0723180101 & 21.8 & 21.8 & 20.8 && 19.8 & 19.6 & 15.9 & \\
		& 0723180201 & 21.6 & 21.5 & 19.9 && 11.9 & 10.6 & 3.0 & \\
		& 0723180301 & 23.9 & 23.8 & 22.7 && 16.3 & 18.1 & 9.3 & \\
		& 0723180401 & 18.6 & 18.8 & 20.7 && 7.1 & 6.6 & 3.3 & \\
		& 0723180601 & 25.6 & 25.5 & 23.9 && 9.1 & 9.8 & 7.0 & \\
		& 0723180701 & 20.9 & 20.8 & 19.2 && 7.3 & 7.5 & 5.5 & \\
		& 0723180801 & 14.7 & 14.6 & 19.0 && 13.6 & 13.6 & 9.7 & \\
		& 0723180901 & 23.2 & 23.1 & 21.5 && 13.2 & 13.7 & 8.9 & 376.7 \\
		\hline
		NGC~3607 	& 0099030101 & 22.3 & 22.3 & 20.0 && 15.6 & 17.6 & 10.8 & \\
		& 0693300101 & 44.4 & 44.4 & 43.8 && 31.3 & 35.3 & 19.8 & 130.4 \\
		\hline
		NGC~3665	& 0052140201 & 40.6 & 40.6 & 36.3 && 27.8 & 29.6 & 20.3 & 77.6 \\
		\hline
		NGC~4382	& 0201670101 & 33.5 & 33.5 & 33.2 && 18.8 & 18.8 & 14.4 & \\
		& 0651910401 & 41.7 & 41.6 & 40.6 && 33.6 & 34.1 & 25.7 & \\
		& 0651910501 & 41.2 & 41.3 & 40.2 && 29.9 & 29.6 & 24.1 & \\
		& 0651910601 & 41.2 & 41.1 & 40.1 && 27.7 & 28.7 & 22.1 & \\
		& 0651910701 & 36.5 & 36.5 & 34.9 && 28.3 & 28.3 & 24.2 & 388.3 \\
		\hline
		NGC~4459	& 0550540101 & 81.8 & 81.8 & 82.8 && 72.7 & 73.2 & 60.6 & \\
		& 0550540201 & 20.4 & 20.4 & 18.8 && 18.8 & 18.8 & 15.1 & 259.1 \\
		\hline
		NGC~4526	& 0205010201 & 25.9 & 25.9 & 26.0 && 22.0 & 21.9 & 17.7 & 61.6 \\
		\hline
		NGC~5353 	& 0041180401 & 22.3 & 22.3 & 20.0 && 21.3 & 21.1 & 16.8 & 59.1 \\
		\hline
	\end{tabular}
	\label{tab:data}
\end{table*}

\begin{figure}
	\centering
	\includegraphics[width=1.02\linewidth]{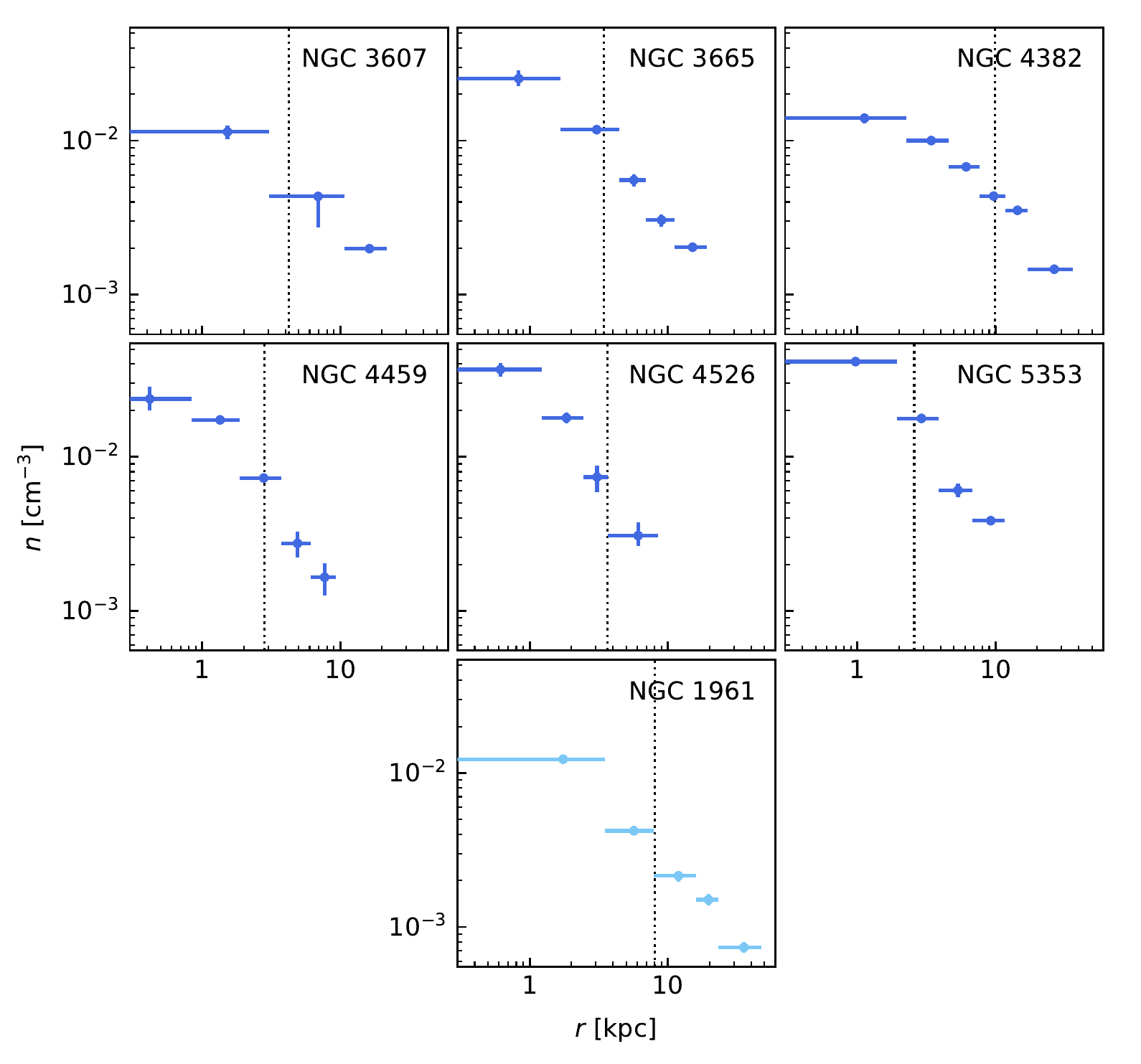}
	\caption{Particle density profiles derived from deprojected spectra.}
	\label{fig:n-single}
\end{figure}

\begin{figure}
	\centering
	\includegraphics[width=1.02\linewidth]{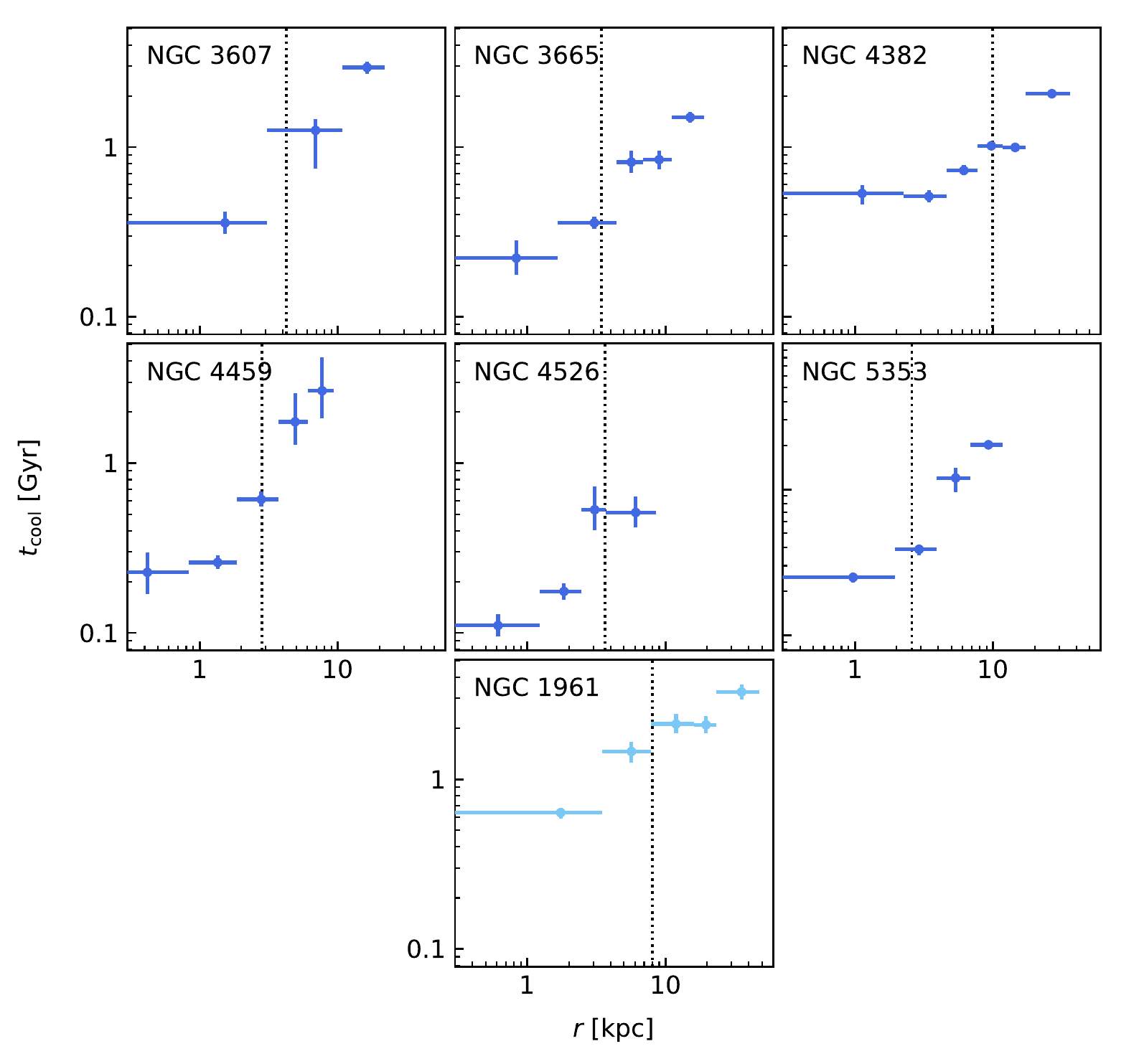}
	\caption{Cooling time profiles derived from deprojected spectra.}
	\label{fig:tc-single}
\end{figure}

\begin{figure}
	\centering
	\includegraphics[width=1.02\linewidth]{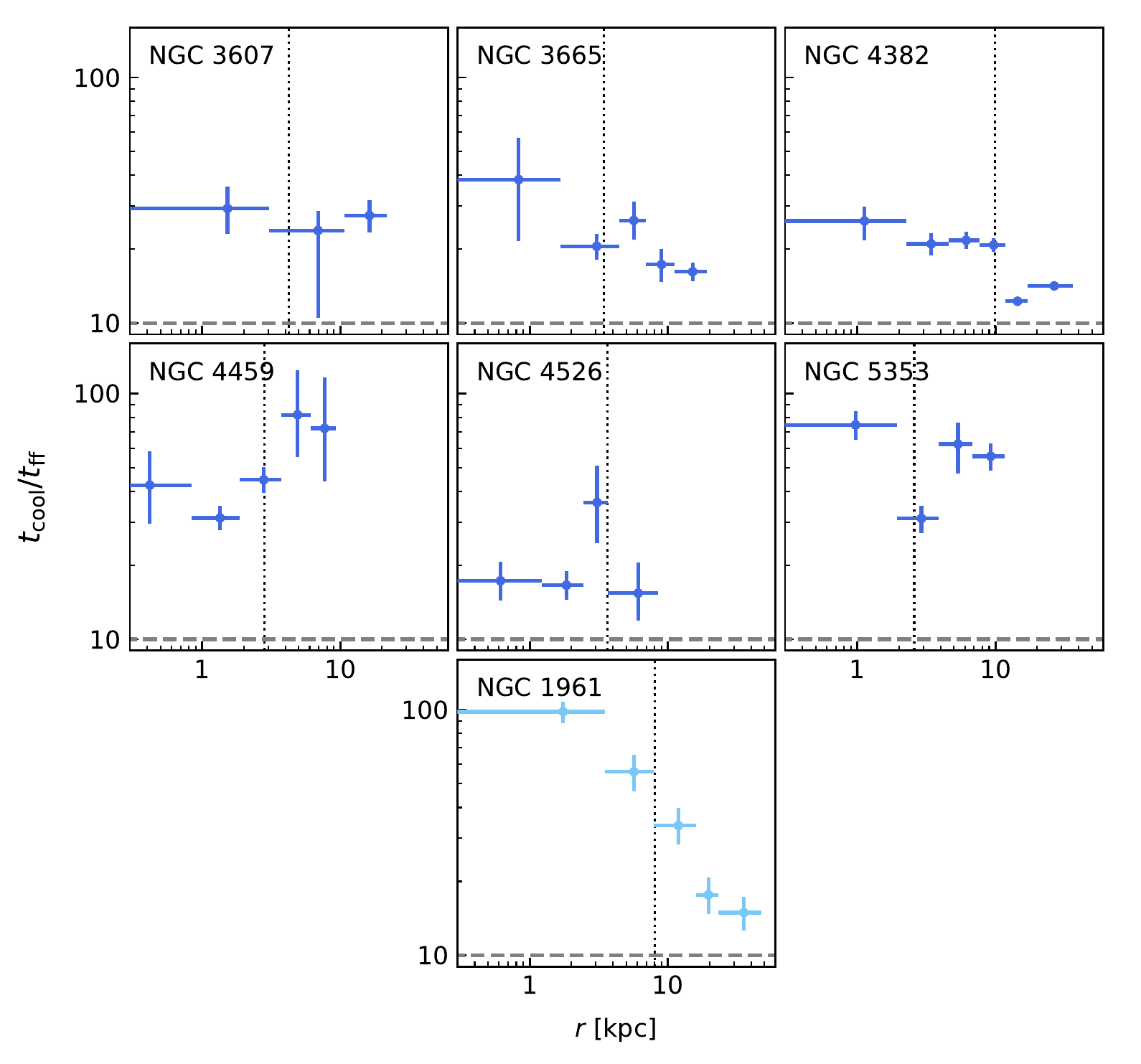}
	\caption{Ratio of cooling time to free-fall time. The value $t_{\rm cool}/t_{\rm ff} = 10$ is visualised as a dashed grey line. }
	\label{fig:tctff-single}
\end{figure}

%%%%%%%%%%%%%%%%%%%%%%%%%%%%%%%%%%%%%%%%%%%%%%%%%%

% Don't change these lines
\bsp	% typesetting comment
\label{lastpage}
\end{document}